\documentclass[11pt, reprint]{revtex4-1}
\usepackage{amsfonts}
\usepackage{amssymb}
\usepackage{amsmath}
\usepackage{bm}
\usepackage{graphicx}
\usepackage{epsfig}

\newcommand{\kk}{{\mathbf k}}
\newcommand{\pp}{{\mathbf p}}
\newcommand{\qq}{{\mathbf q}}
\newcommand{\rr}{{\mathbf r}}

\newcommand{\vv}{{\mathbf v}}

\newcommand{\kp}{p}
\newcommand{\kc}{c}
\newcommand{\si}{{\scriptstyle{K}}}
\newcommand{\sii}{{\scriptstyle{P}}}
\newcommand{\siii}{{\scriptstyle{Q}}}

\begin{document}

\title[Gyrokinetic turbulence]{Gyrokinetic turbulence: between idealized estimates and a detailed analysis of nonlinear energy transfers}
\author{Bogdan Teaca}
\affiliation{Applied Mathematics Research Centre, Coventry University, Coventry CV1 5FB, United Kingdom}
\author{Frank Jenko}
\affiliation{Department of Physics and Astronomy, UCLA, 475 Portola Plaza, Los Angeles, CA 90095-1547, USA}
\author{Daniel Told}
\affiliation{Max-Planck-Institut f\"ur Plasmaphysik, Boltzmannstr. 2, D-85748 Garching, Germany}
\affiliation{Department of Physics and Astronomy, UCLA, 475 Portola Plaza, Los Angeles, CA 90095-1547, USA}
\date{\today}
\begin{abstract}
Using large resolution numerical simulations of gyrokinetic (GK) turbulence, spanning an interval ranging from the end of the fluid scales to the electron gyroradius, we study the energy transfers in the perpendicular direction for a proton-electron plasma in a slab {equilibrium} magnetic geometry. {The plasma parameters employed here are relevant to kinetic Alfv\'en wave turbulence in solar wind conditions.} In addition, {we use an idealised test representation for the energy transfers between two scales, to aid our understanding of the diagnostics applicable to the nonlinear cascade in an infinite inertial range.} For GK turbulence, a detailed analysis of nonlinear energy transfers that account for the separation of energy exchanging scales is performed. 
{Starting from the study of the energy cascade and the scale locality problem, we show that the general nonlocal nature of GK turbulence, captured via locality functions, contains a subset of interactions that are deemed local, are scale invariant (i.e. a sign of asymptotic locality) and possess a locality exponent that can be recovered directly from measurements on the energy cascade. It is the first time that GK turbulence is shown to possess an asymptotic local component, even if the overall locality of interactions is nonlocal.} The results presented here and their implications are discussed from the perspective of previous findings reported in the literature and the idea of universality of GK turbulence. 
\end{abstract}
\maketitle
\onecolumngrid

\section{Introduction}

Plasma turbulence is ubiquitous{, being found} in astrophysical~\cite{Matthaeus:2011p1705} and laboratory~\cite{Boozer:2005p19} settings. Although laboratory experiments~\cite{Gekelman:2011p1328, Theiler:2009p1053} offer a controlled environment, it is often solar wind measurements~\cite{Bruno:2013p1931}  that allow {for plasma turbulence to be probed across a wide range of scales~\cite{Kiyani2015}}. While {the dynamics at large scales} can be captured by fluid approximations~\cite{Goldreich:1997p197, Zhou:2004p21}, the physics of weakly collisional plasma turbulence {requires a kinetic description~\cite{Marsch:2006p1930} for scales comparable to the proton gyroradius and smaller}. Typically, the kinetic scales are associated with the dissipation range of turbulence. In this interval, {which is} important for the plasma heating problem~\cite{Tenbarge:2013p1730, Wan:2015p1754}, {kinetic effects such as} cyclotron~\cite{Lutomirski:1970p1692, Olson:1972p1698} and Landau damping~\cite{Plunk:2013p1890, Klein:2016p1922} need to be considered. However, {a kinetic Alfv\'en wave~\cite{Howes:2008p1524, Howes:2008p1132} and entropy cascades~\cite{Schekochihin:2008p1034, Schekochihin:2009p1131} develop in the same kinetic scales} interval, {through a process similar to the one found in fluid turbulence.} 

{As with neutral fluids}, turbulence in {a} plasma develops due to the existence of couplings between system scales, albeit with a series of complications {resulting from the interactions with the electromagnetic field.} {At kinetic scales these complications are increased even further, as the underlying interactions manifest themselves in a position-velocity phase space.} For a kinetic plasma, the energy redistribution due to nonlinear interactions and the particle-wave resonance (e.g. cyclotron and Landau resonance) cannot be studied separately, as these phenomena influence each other in a turbulence setting~\cite{Plunk:2013p1890, Villani:2014p1545}. For example, in magnetised plasmas an anisotropy develops in the position and velocity space, which influences the balance between the linear phase mixing~\cite{Watanabe:2006p1444, Zocco:2011p1306, PLA:10293237} (that includes linear Landau damping) and nonlinear phase mixing~\cite{Tatsuno:2009p1096, Schekochihin:2008p1034, Hammett:1992p1538, Plunk:2011p1357, Plunk:2010p1360} that occurs in the perpendicular direction and is caused by the same nonlinear term responsible for the generation of the turbulence cascade. Perpendicular plasma structures, generated through nonlinear interactions, can be damped in the parallel direction through Landau damping, if a balance between the damping rates and the nonlinear turnover time can be achieved. While this is a problem that {is receiving more interest within the community~\cite{Hatch:2013p1869, Hatch:2014p1639, Loureiro:2013p1860, Numata:2015p1880, PLA:10293237}} and is far from being solved, analysing the effective perpendicular cascade in {position} space can offer {an} important insight into the properties of kinetic turbulence. 

{Although in kinetic plasma turbulence the electromagnetic fluctuations take part in the particle-wave interactions and mediate the nonlinear couplings, the balance between linear and nonlinear phase mixing can be seen as a species dependent process. It is easy to see that particle-wave resonance conditions depend on the characteristics of the particles. However, the fact that the nonlinear interactions conserve free energy for each plasma species independently, makes the energy cascade a species dependent process as well. Since the properties {of} the nonlinear energy redistribution in kinetic turbulence mirror the ones found in classical fluid turbulence, it is best to investigate the energy cascade for each plasma species independently. This allows us to maximise the applicability of the lessons learned from classical fluid turbulence to our current study of gyrokinetic (GK) turbulence. }

\subsection{A small overview of past studies on energy transfers and locality in turbulence}

The unsolved problem of turbulence has been posed and analysed extensively in the framework of fluid dynamics~\cite{Frisch}. {In} plasmas, the incompressible magnetohydrodynamic (MHD) limit represents the simplest mathematical representation for the turbulence problem. MHD turbulence, considered either for a simple electrically conducting fluid or seen as capturing large scale plasma effects, shares a lot of its properties with its electrically neutral fluid counterpart. While its Alfv\'enic nature, i.e. the existence of Alfv\'en  waves that affect the nonlinear interaction time and the sweeping/straining motions of turbulent structures~\cite{Zhou:2004p21}, does lead to a series of particularities not seen in neutral fluids~\cite{Kraichnan:1965p932, Goldreich:1995p724, Verma:1996p434, Boldyrev:2006p196, Boldyrev:2009p820, Matthaeus:2008p53}, MHD turbulence can still be seen as the nonlinear coupling of scales {which leads} to an intermittent energy cascade.

{In turbulence, the concept of the energy cascade represents the phenomenological interpretation of the effective energy transfer that occurs between any two scales, as a result of the nonlinear interactions~\cite{Kraichnan:1959p1578, Kraichnan:1971p1138}}. For three-dimensional turbulence, the direct energy cascade (i.e. {from large to small scales}) is assumed to be local by Kolmogorov scaling estimates. However, only in the early 90's did numerical simulations allow {the} diagnostics that measure the transfers between two scales to be computed directly from the nonlinear terms, and the direct and local character of the cascade to be shown explicitly. These diagnostics were introduced for neutral fluid turbulence~\cite{Domaradzki:1988p941, Domaradzki:1990p145, Ohkitani:1992p956,Zhou:1993p950}, then ported {to} drift-wave plasma turbulence~\cite{Camargo:1995p1564, Camargo:1996p1563} and latter extensively used for MHD turbulence~\cite{Dar:2001p209, Verma:2004p206, Verma:2005p87, Alexakis:2005p304, Debliquy:2005p203, Carati:2006p632, Alexakis:2007p747, Mininni:2011p1632}. Recently, they were analysed in the context of gyrokinetic {plasma} turbulence~\cite{Tatsuno:2010p1363, BanonNavarro:2011p1274,Nakata:2012p1387}. 

In addition to the {characterisation of the energy cascade as local}, the {problem of locality of nonlinear interactions was studied directly in} fluid~\cite{Kraichnan:1959p1578, Domaradzki:1992p938, Zhou:1993p949, Domaradzki:2007p133, Domaradzki:2009p557, Eyink:1994p856, Eyink:2005p854, Eyink:2009p808, Aluie:2009p809}, MHD~\cite{Aluie:2010p946,Domaradzki:2010p1012, Teaca:2011p1362} and gyrokinetic~\cite{Teaca:2012p1415, Teaca:2014p1571, Told:2015p1712} turbulence. {While the locality of the cascade and the locality of the nonlinear interactions are related, the two concepts possess different characteristics, as we will explore in the present article for gyrokinetic turbulence. Next, we try to clarify the ideas and definitions related to locality in turbulence.}

\subsection{Clarifying the meaning of locality in turbulence}

While not difficult as a generic concept, the various terms used to refer to scale locality in turbulence can become confusing, especially when a detailed analysis is attempted. We clarify what we mean by these terms, which will be used in the current article. 

The {\it locality of the energy exchanging scales} refers to the separation between the energy giving and the energy receiving scales that take part in an energy transfer. This can be evaluated regardless of the information of the mediator scale (i.e. the scale of the advecting field, the third scale involved in a coupling), or by integrating first over all possible mediators. In the latter case, we recover the {\it locality of the energy cascade} problem (referred simply as the {\it locality of the cascade}).

By comparison, the {\it locality of the nonlinear interactions} accounts explicitly for the mediator scale. It relates the intensity of an energetic coupling with the maximal separation that exists between the energy receiving scale and the other two scales. For example, a strong energy transfer between two close scales that is mediated by a much larger scale will contribute to the local character of the cascade, while at the same time it will enhance the nonlocal character of the nonlinear interactions. The locality of the nonlinear interactions is measured by locality functions~\cite{Kraichnan:1959p1578, Teaca:2014p1571} and is traditionally referred to as the {\it locality problem in turbulence} (or simply as the {\it locality problem}). While in both situations we relate the intensity of energy exchanges with the scale separation, the locality of the cascade can be seen as being included in the locality problem.

As nonlinear interactions become scale invariant in the inertial range, i.e. the range where all other interactions are subdominant, the locality problem is also expected to become scale invariant. This implies that the intensity of the energy transfers decreases in the same manner with the increase in separation, regardless of the value of the energy receiving scale (i.e. the reference scale from which we measure the separation in all cases). We now say we recover {\it asymptotic locality}, since once turbulence develops an inertial range the locality problem does not change further. In the inertial range, expressing the decreases in intensity of the energy transfers as a power law of the scale separation yields an exponent, which we call {\it asymptotic locality exponent}.

The asymptotic locality exponent is a characteristic of turbulence. If by varying the parameters that define the system (e.g. plasma parameters) we obtain the same asymptotic locality exponent, then the nonlinear interactions are invariant in regard to these parameters. This invariance of the nonlinear interactions in the inertial range will ensure that any information introduced at large scales (e.g. via large scale forces or boundaries) will be destroyed (decorrelated) through the cascade process. At the end of the cascade the same information is recovered, which leads to the small scales to be universal. Traditionally this is referred to as the {\it universality of turbulence} problem, since universal small scales lead in classical fluids to universal dissipative mechanisms. However, we can allow the small scales to be directly affected by some linear mechanism, while at the same time having a universal character for the nonlinear interactions. Defining the universality of turbulence as the universality of nonlinear interactions is appropriate, as it is the latter we need to have to be able to develop unique sub-grid scale models. Regardless of the accepted definition of universality, a unique asymptotic locality exponent is seen as a necessary condition for the existence of universality in turbulence.

\subsection{Structure of the article}

In the current paper, we study the effective energy transfer in the perpendicular direction for a magnetised plasma described by a gyrokinetic (GK) formalism. We are interested in describing the energy cascade and {the locality problem for} GK turbulence. This analysis relates to the fundamental question of universality of turbulence and in particular the universality of plasma turbulence. {Starting from the study of the energy redistribution and the scale locality problem, we show that the general nonlocal nature of GK turbulence, captured via locality functions, contains a subset of interactions that are deemed local, are scale invariant (i.e. a sign of asymptotic locality) and possess an exponent that can be recovered directly from the interaction of solely energy exchanging scales (i.e. the energy cascade). }

In section~\ref{sec:GK}, we present the GK equations for a slab magnetic geometry and list the parameters of the nonlinear simulation employed throughout this work. In addition, we derive the free energy balance equation for a scale and define a norm for the intensity of the energy transfers in the system. In section~\ref{sec:scales}, we {discuss succinctly} the magnetic geometry effects on the scale representation, present the interaction conditions between three scales and the resulting implications on their separation and introduce the waveband decomposition for the GK system. In section~\ref{sec:interactions}, after building the {nonlinear energy transfers} and the scale flux diagnostics, listing their properties and interconnections, we proceed to present the results for the numerical simulation employed. This section presents diagnostics which can be seen up to a point as classical tools for the analysis of turbulence, i.e. tools used in the past for the analysis of the equivalent problem in fluids.  

In an attempt to understand better the connection between energy transfers and the locality of interactions, we introduce in section~\ref{sec:test} an idealised test problem. In section~\ref{sec:measurements}, employing the lessons learned from the test problem and applying a series of detailed considerations, we conduct a further analysis of the large resolution GK simulation data. We find that the locality exponents for the energy cascade exhibit an asymptotic behavior, denoting the possibility of a universal character for the energy cascade in GK turbulence. Furthermore, we show that these exponents can be obtained directly from the energy transfers between two scales rather than the computationally intensive locality functions (modified in section~\ref{sec:measurements} to account only for the locality of the energy cascade). These implications and their link with past results presented in the literature are discussed last, in section~\ref{sec:conclusions}.

\section{The gyrokinetic framework} \label{sec:GK}

The gyrokinetic formalism represents a rigorous limit \cite{Brizard:2007p11} of kinetic theory for strongly magnetised plasmas for which gyrotropy is assumed (i.e. invariance under the gyration of particles of charge $q_\sigma$ and mass $m_\sigma$ around the magnetic guide field of intensity $B_0$, for each plasma species $\sigma$). We reproduce below the main ideas behind GK for the simpler context of astrophysical plasmas~\cite{Howes:2006p1280}.

At the kinetic level, the distribution functions expressed at the particle position $f_\sigma(\rr,\vv,t)$ represent the dynamical quantities of interest.  The role of the self-consistent electromagnetic fields, obtained from {velocity} moments of the particles' distributions, is to mediate the linear and nonlinear interactions between structures in the distribution functions. The most common approach for kinetic turbulence is to assume small fluctuations around equilibrium background distribution functions $F_\sigma$, considered usually to be Maxwellians. {In the GK formalism, the dynamics are contained by the perturbed gyro-center distribution functions} $h_\sigma(x,y,z, \mu, v_\parallel,t)$, where $(x,y,z)$ are magnetic coordinates, {with the $z$ direction coinciding with the direction of the guiding magnetic field lines}. Aligning the representation of the system with the direction of homogeneity induced by gyrotropy allows us to remove the gyration motion (of gyroradius $\rho_\sigma= \sqrt{T_{\sigma} m_{\sigma}} c / e B$) from the kinetic system and effectively reduce the phase space to just five-dimensions (i.e. $x,y,z, \mu, v_\parallel$). This reduction is important for the numerical implementations~{\cite{Garbet:2010p1103}}, as it substantially saves computational resources by considering only two velocity directions. The velocity along the magnetic field line is $v_\parallel$. {The magnetic moment ($\mu=m_\sigma v^2_\perp/2B_0$) is an adiabatic invariant for the GK system and contains implicitly the perpendicular velocity $v_\perp$ information}. Typically, the velocity directions are expressed in thermal velocity units ($v^{th}_\sigma=\sqrt{2T_\sigma/m_\sigma}$) and the equilibrium (background) density $n_{\sigma}$ and temperature $T_{\sigma}$ are known. 

Accounting for a Boltzmann response factor due to the process of restoring plasma electroneutrality and writing explicitly only terms up to the first order in the small parameter introduced by the GK ordering (notably low frequencies for the plasma fluctuations compared to the ion, here proton, cyclotron frequency $\Omega_\sigma$ and small fluctuation levels), we obtain \cite{Howes:2006p1280} a relation between $f_\sigma(\rr,\vv,t)$ and the perturbed gyro-center distribution functions $h_\sigma(x,y,z, \mu, v_\parallel,t)$,
\begin{align}
f_\sigma=F_\sigma [1- \frac{q_\sigma\phi}{T_\sigma} \big{]}+h_\sigma \ .
\end{align}
For such an expansion to be valid and for the removal of the particles' fast gyro-motions to be done systematically, the GK ordering must hold from a physics perspective. {In strongly magnetised plasmas, the GK formalism represents a rigorous kinetic representation of the turbulence problem, being able to capture the kinetic Alfv\'en wave cascade~\cite{Howes:2011p1370, Howes:2011p1459}, as well as the entropy cascade~\cite{Schekochihin:2008p1034, Schekochihin:2009p1131} and the linear phase mixing associated with Landau damping. In astrophysical studies,} while it neglects cyclotron resonance, gyrokinetics {can still be seen as a useful tool as it captures \cite{Told:2016p1915} the crucial dynamics of  kinetic Alfv\'en wave (KAW) turbulence in three spatial dimensions~\cite{podestasp13} while offering a} more manageable system to be simulated numerically. The GK equation is just the Vlasov equation rewritten for $h_\sigma(x,y,z, \mu, v_\parallel,t)$ and considering gyrotropy, for which the electromagnetic potentials are determined from electromagnetic sources obtained at the location of the particles' gyroradius (i.e. the electromagnetic sources are effectively rings of electric charge centered on the gyro-centers).

\subsection{The gyrokinetic equations}

The nonlinear gyrokinetic equations~\cite{Frieman:1982p1941} were derived rigorously by Ref.~\cite{Hahm:1988p1943} and presented extensively in~\cite{Brizard:2007p11}. An appropriate review of GK turbulence for newcomers is {presented} by Ref.~\cite{Krommes:2012p1373} and in the simplifying context of astrophysical plasma the GK formalism is presented in~\cite{Howes:2006p1280}. In slab magnetic geometry, for $h_\sigma=h_\sigma(x,y,z,v_\parallel, \mu)$ the gyro-center distribution functions, the gyrokinetic equation for a species $\sigma$ has the form
\begin{align}
\frac{\partial h_\sigma}{\partial t}= \underbrace{ -[\frac{c}{B_0} {\bf e}_z \times \nabla \bar \chi_\sigma]}_{{\bf v}_{\sigma}} \cdot \nabla h_\sigma - v_{\parallel} \frac{\partial h_\sigma}{\partial z}+ \frac{q_\sigma F_{\sigma}}{T_{\sigma}}\frac{\partial \bar \chi_\sigma}{\partial t} -  \left[\frac{\partial h_\sigma}{\partial t}\right]_{coll}   \label{GKeq}
\end{align}
where $\displaystyle \bar \chi_\sigma= \bar \phi_\sigma-\frac{v_\parallel}{c} \bar A_{\parallel,\sigma} + \frac{\mu }{q_\sigma}\bar B_{\parallel,\sigma} $ is the gyrokinetic (gyro-averaged) potential and $ \phi$, $A_\parallel$ and $B_\parallel$ are {obtained \cite{Gorler:2011p1356, Gorler:2011p1340} from their respective field equations 
\begin{align}
\nabla^2_\perp \phi &= -4\pi \sum_\sigma q_\sigma  n_\sigma  \\
\nabla^2_\perp A_{\|} &= -\frac{4\pi}{c} \sum_\sigma j_{\|,\sigma} \\
B_{\|} &= - \frac{4\pi}{B_0} \sum_\sigma  p_{\perp,\sigma}
\end{align}
for known sources ($n_\sigma$, $j_{\|,\sigma}$, $p_{\perp,\sigma}$), which in turn are obtained from the {velocity} moments of $h_\sigma$ at the location of the particles' gyroradius. As this is expressed in a simpler form in $\kk_\perp$ space, we list the sources equations for a mode $\kk_\perp$ as
\begin{align}
n_\sigma(\kk_\perp)&= \frac{2\pi B_0}{m_\sigma} \int\!\!d v_\| d\mu \,\, \left[J_0(a_\sigma) h_{\sigma}(\kk_\perp)  - q_\sigma \phi_\sigma(\kk_\perp)  \frac{F_{\sigma}}{T_{\sigma}}\right]  \\
j_{\|,\sigma}(\kk_\perp)&= q_\sigma \frac{2\pi B_0}{m_\sigma} \int\!\!d v_\| d\mu \,\, v_\| \left[ J_0(a_\sigma) h_{\sigma}(\kk_\perp) - q_\sigma \phi(\kk_\perp)  \frac{F_{\sigma}}{T_{\sigma}} \right]  \\
p_{\perp,\sigma} (\kk_\perp)&= \frac{2\pi B_0}{m_\sigma} \int\!\!d v_\| d\mu \,\, \mu B_0 I_1(b_\sigma) h_{\sigma}(\kk_\perp) 
\end{align}
where $J_0$ is the Bessel function, $I_1$ is the the modified Bessel function and for $k_\perp\equiv|\kk_\perp|$ we have $a_\sigma=k_\perp \sqrt{\frac{2B_0 \mu}{m_\sigma \Omega_\sigma}}$ and $b_\sigma=\frac{1}{2}({v^T_\sigma k_\perp/ \Omega_\sigma})^2$~\cite{Morel:2011p1339}.} The term $\left[\frac{\partial h_\sigma}{\partial t}\right]_{coll}$ refers to the impact made on the time evolution of $h_\sigma$ by a linearised Landau-Boltzmann collision operator {(see the supplemental material provided by Ref.~\cite{Navarro:2016p1965} for the exact form used here).} {While not standard, we absorb a minus sign in the definition ${\bf v}_{\sigma}$, which is minus the drift velocity, as a way to achieve a more compact notation in the next sections.}

\subsection{Nonlinear simulation data} \label{sec:GK:data}

In this study we use gyrokinetic simulations of magnetised proton-electron plasmas. The nonlinear gyrokinetic system of equations is solved {using} the Eulerian code {\sc GENE}~\cite{gene}. The data used in this work is taken from the simulation presented in Ref.~\cite{Told:2015p1712}, one of the largest GK simulation to date, and it is briefly summarised below. {We mention that, while our analysis is limited to the use of this pre-existing large resolution simulation, and an even larger velocity space domain and velocity space resolution would be needed ideally for diagnosing the overall phase space mixing problem, we are confident in the results presented here, results which relate to the energy cascade in the perpendicular spatial direction and the locality of interactions problem.}

The physical parameters of the simulation {are} chosen to be close to the solar wind conditions at 1 AU, with $\beta_{i}=8\pi n_{i}T_{i}/B_{0}^{2}=1$ and $T_{i}/T_{e}=1$. Proton and electron species are included with their real mass ratio of $m_{i}/m_{e}=1836$. The electron collisionality is chosen to be $\nu_{e}=0.06\, \omega_{A0}$ (with $\nu_{i}=\sqrt{m_{e}/m_{i}}\nu_{e}$), and $\omega_{A0}$ being the frequency of the slowest Alfv\'en wave in the system.  The evolution of the gyro-center distribution is tracked on a grid with the resolution $\{N_{x}, N_{y}, N_{z}, N_{v_{\parallel}}, N_{\mu},N_{\sigma}\}=\{768, 768, 96, 48, 15, 2\}$, where ($N_x,N_y$) are the perpendicular, $(N_z)$ parallel, $(N_{v_{\parallel}})$ parallel velocity, and $(N_{\mu})$ magnetic moment ($\mu\!=\!m_\sigma v^2_\perp/2B_0$) grid points, respectively. This covers a perpendicular dealiased wavenumber range of $0.2\le k_{\perp}\rho_{i}\leq 51.2$ (or $0.0047\leq k_{\perp}\rho_{e}\leq1.19)$ in a domain $L_x=L_y=10\pi\rho_i$. In the parallel direction, a $L_z=2\pi L_{\parallel}$ domain is used, where $L_{\parallel} \gg \rho_i$ is assumed by the construction of gyrokinetic theory. A velocity domain up to three thermal velocity units is taken in each direction. The fluctuations in the system are driven via a magnetic antenna potential (a $A_\parallel^{\scriptsize \mbox{ant}}$ contribution is added to $\chi$), which is prescribed solely at the largest scale and evolved in time according to a Langevin equation~\cite{tenBargecpc14}. We mention that this antenna potential is removed from $\bar \chi_\sigma$ before the nonlinear diagnostics are computed.

\subsection{The free energy balance equation for a scale}

The free energy for a species ($\mathcal{E}_\sigma$) represents the quadratic quantity of interest for the study of {the dynamics of} gyrokinetic turbulence \cite{Schekochihin:2009p1131}. Free energy is the quantity that is injected into the system by instabilities or external drives, dissipated by collisions, while being redistributed in a conservative fashion by the action of the nonlinear terms ($\mathcal{E}_\sigma$ is nonlinear conserved for each species $\sigma$ independently). The free energy for a species $\sigma$ is defined using the GK system as
\begin{align}
\mathcal{E}_\sigma  = \left<\frac{T_\sigma}{2 F_\sigma}h_\sigma\left[ h_\sigma - \frac{qF_\sigma}{T_\sigma}\bar \chi_\sigma  \right]   \right>  \;, 
\label{fedef}
\end{align}
where the $\left<\cdots \right>$ notation stands for the average over the phase space domain, including the appropriate {five-dimensional Jacobian ($J_{5D}$)} contributions,
\begin{align}
\left<\cdots \right>= \bigg{[}\int (\cdots)J_{5D}\, dx\, dy\, dz\, dv_\parallel d\mu \bigg{]} \bigg{/} \bigg{[}\int J_{5D} \, dx\, dy\, dz\,  dv_\parallel d\mu \bigg{]} \;. 
\end{align}
See Ref.~\cite{ABNThesis} appendix B for a full derivation of the free energy definition for GK, starting from its classical form (entropy contribution + electric energy + magnetic energy). While the link with eq.~(\ref{fedef}) is conceptually straightforward, the derivation is too tedious to be reproduced here and we consider the free energy definition for GK theory to be granted by eq.~(\ref{fedef}).

Rewriting eq.~(\ref{GKeq}) as
\begin{align}
\frac{\partial }{\partial t}  \left[ h_\sigma - \frac{q_\sigma F_\sigma}{T_\sigma}\bar \chi_\sigma  \right]= {\bf v}_\sigma \cdot \nabla h_\sigma - v_{\parallel} \frac{\partial h_\sigma}{\partial z}  -  \left[\frac{\partial h_\sigma}{\partial t}\right]_{coll}  \label{GKtwo}\;,
\end{align}
we can formally obtain the global free energy balance equation for the species $\sigma$ by applying the operator $\big{<}\frac{T}{2 F}h_\sigma \cdots \big{>}$ on each of the terms in eq.~(\ref{GKtwo}). Since the free energy is a nonlinear invariant, we have $\big{<}\frac{T}{2 F}h_\sigma ({\bf v}_\sigma \cdot \nabla h_\sigma) \big{>}=0$. A much more useful balance equation is obtained for a hierarchy of scales, naturally provided by a Fourier decomposition, by considering
\begin{align}
h_\sigma^{[k]}(x, y, z, v_{\parallel}, \mu)= \!\!\! \int\displaylimits_{ |{\bf k}_\perp|= k}   \!\!\!\!\!\!\widehat h_\sigma(k_x, k_y, z, v_{\parallel}, \mu) e^{i(k_x x+k_y y)}dk_x dk_y \;.
\label{hdk}
\end{align}
The free energy balance equation for a scale is now simply obtained by applying the filtered operator $\big{<}\frac{T}{2 F}h_\sigma^{[k]} \cdots \big{>}$ on each of the terms in eq.~(\ref{GKtwo}) and has the form
\begin{align}
\frac{\partial \mathcal{E}_\sigma(k)}{\partial t} = \mathcal{T}_\sigma(k) + \mathcal{L}_\sigma(k) + \mathcal{D}_\sigma(k)   \label{GK:FE}\;.
\end{align}
The $\mathcal T_\sigma(k)$ contains the contribution of all nonlinear interactions for a scale identified by $k$ and is known as the {\em transfer {spectrum}}. The $\mathcal{L}_\sigma(k)$ term contains the linear parallel dynamics (including linear Landau damping). {As long as the filtering condition does not depend on $z$ (e.g. $dk/dz=0$, the case considered here), $\mathcal{L}_\sigma(k)$ is zero for all $k$'s due to the $\{z, v_\parallel \}$ integration. Otherwise, the geometry leads to a (linear) $k$-redistribution of energy that integrates to zero over all $k$-scales.} Last, $\mathcal{D}_\sigma(k)$ is a dissipative term that appears due to the presence of collisions.

\subsection{Choosing a norm for the energy transfers}

The {transfer spectrum} $\mathcal T_\sigma(k)$ integrates to zero over all scales, a result of the conservation of free energy by the nonlinear interactions. It represents the simplest quantity related to the nonlinear energy exchanges that can be computed numerically and it can be recovered from all other nonlinear diagnostics that account for additional scale decompositions of the nonlinear term, as we will see in later sections. Taking all these facts into account, we consider that $\mathcal T_\sigma(k)$ can serve as a basis for a useful norm that will allow us to gauge the intensity of various energy transfers. We formally define this norm $\varepsilon_\sigma$ to be:
\begin{align}
\varepsilon_\sigma=\frac{1}{2}\int \big{|} \mathcal T_\sigma(k)\big{|} dk\, .
\end{align}

{In general,} this is a definition that requires no particular shape for {a} $\mathcal T(k)$ curve, except that it integrates to zero over the entire scale domain considered. {For steady state turbulence that exhibits a clear inertial range (i.e. the value of $\mathcal T(k)$ goes from negative to positive with the increase in $k$ and is zero in the inertial range), $\varepsilon$ defined above recovers the scale-invariant value for the energy flux in the inertial range. However, the use of this definition is also appropriate for transient state turbulence, for which an inertial range is not observed and a representative energy flux value cannot be determined for the system.} Having units of power, $\varepsilon$ can act as an indicator of the intensity of the nonlinear energy exchanges present in the system {(here for each species $\sigma$) and it can be used} to compare turbulence between various states and simulations.

\section{The representation of perpendicular scales} \label{sec:scales}

Since the gyrokinetic formalism is strongly anisotropic and assumes by construction that $k_\perp \gg k_\parallel$, the main effect of the nonlinear interactions is to mix the perpendicular spatial scales. While we recall that the underlying GK dynamics occur in a five-dimensional phase space and the nonlinear phase mixing does become important at scales $k_\perp \rho_\sigma >1$, the effective energetic interaction between three perpendicular modes ($\kk_\perp+\pp_\perp+\qq_\perp=\bf{0}$) can still be measured and the resulting perpendicular scale interactions can be analysed. To simplify the notations, we use $\kk$ from now on to refer to the perpendicular wave vector $\kk_\perp$. Considering that a scale $\ell$ can be defined by the norm of a wave vector (e.g.  $\ell \sim 1/k$), we will typically identify a scale by the norm $k$ and, by abuse of language, refer to it as a $k$-scale, even though the $k$ norm has units of inverse length.

\subsection{The impact of the magnetic geometry on the perpendicular scales}
\begin{figure}[b]
\center
\includegraphics[width = 0.8\textwidth]{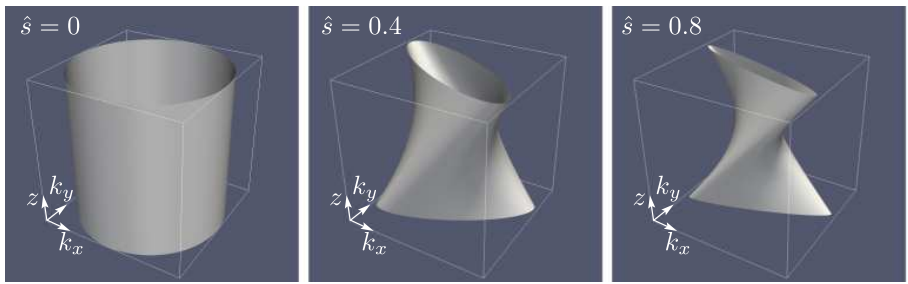}
\caption{The iso-surface of $k(z)$ for $z\in[-\pi,+\pi]$ and for the $k_x, k_y$ domains centered on zero, respectively. The slab case corresponds to $\hat s =0$. Note that the largest $k$ value that is fully captured in the same ($k_x, k_y, z$) domain, as considered in all the figure's panels, decreases with the increase of the shear $\hat s$, as evident from the $z$ mid-plane for which $k(z=0)$ is a circle of diminishing radius. }
\label{FigPerpS}
\end{figure}

Before analysing the interactions between perpendicular scales, we stop to talk briefly about the impact made by the geometry of the magnetic guide field on the scale representation. In the $(\kk, z)$ space, let us consider the magnetic geometry to be prescribed via a contra-variant metric tensor that varies only along the magnetic field, i.e. $\eta^{ij}=\eta^{ij}(z)$. This is one of the simplest scenarios, that of the local approximation of the magnetic flux surfaces typically used in tokamak studies \cite{Lapillonne:2009p1355}. The perpendicular scale, considered as the wave norm $k\equiv|\kk|$, is found as
\begin{align}
k(z)=[\eta^{xx}(z)k_x^2 + \eta^{xy}(z)k_x k_y + \eta^{yy}(z)k_y^2]^{1/2} \label{perpscale}.
\end{align}
We immediately notice that the perpendicular scales have now a $z$ dependence. This simple fact complicates the scale decomposition typically employed in turbulence studies, which now needs to be done in $(\kk,z)$ rather than simply in $\kk$. While in a slab geometry, i.e. the magnetic configuration of choice for astrophysical studies, $\eta^{xx}(z)=\eta^{yy}(z)=1$, $\eta^{xy}(z)=0$ and $dk/dz=0$, let us visualised in figure~\ref{FigPerpS} the $k(z)$ norm for a sheared box given by $\eta^{xx}(z)=1$, $\eta^{xy}(z)=\hat s z$, $\eta^{yy}(z)=1+(\hat s z)^2$ which recovers the slab configuration for $\hat s =0$.

The $z$ dependence for a scale is not just a nuisance that complicates the scale decomposition. The nonlinear interactions occur between a triad of resonant modes ($\kk+\pp+\qq=\bf{0}$). In the slab case the interaction of any three modes leads to an interaction {solely between} three perpendicular scales. {Once} magnetic curvature effects are considered the same triad interaction now couples multiple scales together, as the same $k_x,k_y$ mode contributes to multiple $k(z)$ scales. {Furthermore, in the free energy balance equation (eq.~\ref{GK:FE}) we now have $\mathcal{L}(k) \ne 0$ (without going into details, this can be seen as a variation in terms of $z$ leading to a variation in terms of $k$)}. This shows a split between the magnetic coordinates used to describe the GK system (i.e. $k_x, k_y, z$) and the natural coordinates used for the description of scales in turbulence (i.e. $k$); a nontrivial problem that requires further analysis from the plasma scientific community. This is the main reason why the slab configuration, which we resume to from this point on, is preferred as a basis for understanding the basic dynamics of gyrokinetic turbulence, even though the general gyrokinetic theory can account for magnetic geometry effects. {However, fully understanding the fundamental characteristics of plasma turbulence in arbitrary magnetic geometry is a highly desirable scientific proposition.}

\subsection{Interaction conditions for scales} \label{sec:scalescond}

\begin{figure}[hbt]
\center
\includegraphics[width = 0.8\textwidth]{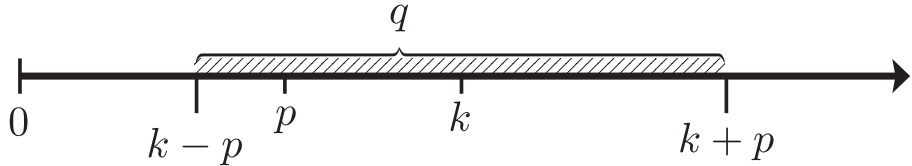}
\caption{The possible range for a scale $q$ that can participate in the interaction of two given scales $p$ and $k$.}
\label{FigScaleDomain}
\end{figure}

As mentioned, the nonlinear energetic interactions occur due to all possible triads, i.e. three modes for which their wave vectors obey the triad condition 
\begin{align}
\kk+\pp+\qq={\bf 0} \label{resonant}.
\end{align}
This triad condition imposes a limit on the interaction of scales.  Using the fact that $|\kk+\pp+\qq|= 0$, we have {the triangle inequalities}:
\begin{align}
&q =|\kk+\pp| \le k+p, \\
&p =|\kk+\qq| \le k+q \Rightarrow q\ge p-k, \\
&k =|\pp+\qq| \le p+q \Rightarrow q\ge k-p. \label{qp}
\end{align}
These conditions tell us that for given $k$ and $p$ scales, the third scale $q$ that enters the nonlinear interaction, respecting the triad condition given by eq. (\ref{resonant}), needs to obey: 
\begin{align}
&q \le k+p, \\
&q \ge \max\{k,p\}-\min\{k,p\} \label{qmin}. 
\end{align}
An example is visually represented in figure \ref{FigScaleDomain} {and more detailed pictures of allowed scale interactions can be found in Ref~\cite{Gurcan:2016p1969}}. We see that for $k \sim p$ the scale $q$ can be at most $2 k$ and at least $0$. A further consideration on scale separation is undertaken next. 

\subsection{Dyadic separation of scales} \label{sec:dyad}

{Let us first consider $q\le k$, $p\le k$ and $k$ as three scales coupled by a nonlinear interaction, where $k$ denotes the smallest scale initially available in the system. The nonlinear interaction in question can potentially generate scales smaller than $k$. From eq.~(\ref{qp}) we see that any new smaller scale is contained in the interval $[k, 2k]$, i.e. at most a factor of $2$ smaller. A phenomenological interpretation can be given: the shearing of a structure by a larger advecting flow can generate at most scales half the structure's size.

We will refer to the interval $[k, 2k]$ as a {\it dyadic band}~\cite{Eyink:1994p856}, and refer to a separation between $p=k/2$ and $k$ as a {\it dyadic separation}. In music, a dyadic band represents an octave.} Let us now consider a dyadic {(octave)} separation of scales (i.e. $...k/4, k/2, k, 2k, 4k...$) and assess the implications for the interaction of two scales thus separated. The interaction of $k$ with any $p\le k/2$ is mediated by the scales $q\in[k/2, 3k/2]$, {according to the triangle inequalities introduced above (section~\ref{sec:scalescond}).} This means that only two dyadic bands, one on each side of $k$ captures the entire interaction process {(that is, two dyadic separated scales can be seen as interacting rather locally)}. The scale $q$ can be less than $k/2$ and tending towards zero {(thus increasing the nonlocal character of the interaction)} only when scale $p$ is less than a dyadic separation away from $k$ (i.e. $k/2 < p < k$). {For a given $k$ and $k/2 < p < k$, we designate their coupling as an {\it {intra-dyadic}} interaction. We also notice that for the separation from $k$ of both $q<k$ and $p<k$ to be equal, it is required that $q=p=k/2$, i.e. both $q$ and $p$ are dyadic separated from $k$. In later sections, these definitions will help us in the classification of local and nonlocal interactions.}

Far from being a simple thought experiment, physical scales {do} need to be {separated sufficiently} (in a $\log(k)$ space) {to be distinguished from each other} (e.g. the octave separation of frequencies for musical notes being a classical example in this sense). {Thus, a flow structure with scales contained by a wavenumber band can be better associated with the concept of an eddy in turbulence, i.e. a structure well-localised at the same time in real space and in wave space, opposed to the case of a simple linear wave that is identified by a single wavenumber and which is completely un-localised in real space. While a dyadic band decomposition can ultimately be seen as arbitrary}, the interaction between two dyadic separated scales can prove to capture better the {phenomenological} interaction between two scale structures in turbulence \cite{Aluie:2010p946} {and provide a simpler link with the classical phenomenological interpretation of turbulence.} Next, we introduce the waveband decomposition for the  scales {that we will actually use in the measurement of the energy transfers.}

\subsection{Waveband representation} \label{sec:scales:waveband}

As in previous works for GK turbulence~\cite{BanonNavarro:2011p1274, Teaca:2012p1415, Teaca:2014p1571,Told:2015p1712}, we define a series of scale intervals $s_n=[k_{n-1},k_n]$, with boundary wavenumbers given as a geometric progression, 

\begin{align}
k_n=k_{1} \lambda^{n-1} \;,
\end{align}
for $n\in \mathbb{N}^*$, $\lambda>1$ and $k_0=0$. These structures are called shells in previous works~\cite{Dar:2001p209, Verma:2004p206, Verma:2005p87, Alexakis:2005p304, Debliquy:2005p203, Carati:2006p632, Alexakis:2007p747, Mininni:2011p1632} (here having the geometric shape of cylindrical shells in a $(\kk, z)$ space) or bands~\cite{Domaradzki:1988p941, Domaradzki:1992p938, Domaradzki:2007p134, Domaradzki:2009p557, Aluie:2010p946}, to account for the fact that they represent bands of equal width in a $\log(k)$ space. 

The distribution functions or electromagnetic potentials are filtered in wave space, obtaining their respective band (shell) filtered contributions. For example, the waveband filtered distribution functions $\widehat h^{[n]}(\kk)$ are found in $\kk$ space as
\begin{eqnarray}
\widehat h^{[n]}(\kk)=\left\{ \begin{array}{lcl}
\widehat h(\kk) , &  |\kk| \in s_n \\
0 , &  |\kk| \notin s_n  
\end{array}  \right.  \;, \label{shell-decomposition}
\end{eqnarray}
while the real space contributions are simply obtained as
\begin{align}
h^{[n]}(x,y)=  \int \widehat h^{[n]}(\kk) e^{i(k_x x+k_y y)}d{\bf k} \;.
\label{deltaH}
\end{align}
It is important to realise that the filtered signals are well defined in real space, the total information being recovered as the superposition of all filtered contributions, e.g. 
\begin{align}
h(x,y)=\sum_{n} h^{[n]}(x,y)\;,
\end{align}
and that they are orthogonal to each other, i.e. $\int h^{[n]}(x,y)h^{[m]}(x,y) dx dy=0$ for $n\ne m$. We mention that {a decomposition using infinitesimally thick bands} could be performed, equivalent to the recovery of the wave-norm $k$-scale splitting prescribed by eq.~(\ref{hdk}). However, a geometric progression is preferred for turbulence studies, since scaling laws play an important part and, we want to separate physical structures without wasting numerical resources.

For the current GK study we take a total of $N=25$ wavebands, with $k_{1}=0.275 \rho_i$ and $\lambda=2^{1/3}$. While a dyadic separation (i.e. $\lambda=2$) is most useful {for the analysis of energy transfers between structures with a more robust phenomenological equivalence, i.e. eddies}, our choice for the $\lambda$ factor allows us to perform a finer analysis of the nonlinear interactions{via the analysis of scale fluxes. Past studies used $2^{1/4}$ (Ref.~~\cite{Teaca:2011p1362}) and $2^{1/5}$ (Ref.~~\cite{Teaca:2012p1415}) as the values for $\lambda$. In the current work, which makes use of Ref.~\cite{Told:2015p1712} large scale GK computations, the $\lambda=2^{1/3}$ choice was done to reduce the number of bands required to spawn the $k$ interval. This is a choice dictated by computational costs, as the most complex diagnostic requires 
the calculations of the nonlinear term $N^2$ times.}

\section{Nonlinear energetic interactions in GK turbulence} \label{sec:interactions}

{In the current work,} instead of using the triad transfers {(i.e. the nonlinear energy transfers that occur between three modes which respect the $\kk+\pp+\qq=0$ resonant condition)} as the basis for the {conceptual} definition of various energy transfers~\cite{Teaca:2012p1415, Teaca:2014p1571,Told:2015p1712}, we employ an alternative
presentation. While the two approaches are in fact equivalent, we consider the following introduction of the transfer of energy between three scales to be easier to grasp.

\subsection{Building the triple-scale transfer}

Suppressing the plasma species index and using for each species the implicit form of the {advecting} velocity, i.e. $\vv\!=\!-\!\left[ \frac{c }{B_0}{\bf e}_z \!\times\! \nabla \bar \chi \right]$, the nonlinear term entering the GK equation has the compact expression $\vv \cdot \nabla h$. The global variation of the species free energy due to the nonlinear term has now the form
\begin{align}
\frac{\partial {\mathcal E}}{\partial t} \bigg{|}_{nonlinear}=\bigg{\langle} \frac{T}{2F}(\vv \cdot \nabla h) h \bigg{\rangle}=0
\end{align}
and is zero due to the conservation of free energy by the nonlinear interactions. Considering the scale decomposition $h\!\!=\!\!\sum_\si h^{[\si]}$, we can rewrite the $\big{\langle} \frac{T}{2F}(\vv \cdot \nabla h) h \big{\rangle}$ term as $\sum_\si \big{\langle} \frac{T}{2F}(\vv \cdot \nabla h) h^{[\si]} \big{\rangle}$. Performing a similar scale splitting on $\nabla h$ and $\vv$ we obtain the equivalent statement for the conservation of free energy by the nonlinear interactions,
\begin{align}
\sum_\siii \sum_\sii \sum_\si \bigg{\langle} \frac{T}{2F}(\vv^{[\siii]} \cdot \nabla h^{[\sii]}) h^{[\si]} \bigg{\rangle}=0 \;.
\end{align}

At this point we make two remarks. First, while globally the energy transfers integrate (sum) to zero, the individual transfers $\big{\langle} \frac{T}{2F}(\vv^{[\siii]} \cdot \nabla h^{[\sii]}) h^{[\si]} \big{\rangle}$ can have any value and will form the basis for our triple-scale transfers. And second, we see that the splitting employed is true for any decomposition of our quantities, not just for a waveband decomposition. In fact, it is up to us to provide the proper physical scale decomposition and justify what can be seen as an arbitrary choice. {We remind the reader that $\si, \sii, \siii$ are integers that identify the waveband intervals and are not themselves a wavenumber, e.g. $\si$ is the integer that identifies the band spanning the $[k_{\si-1}, k_\si]$ interval.} 

Employing a waveband scale decomposition prescribed by eq.~(\ref{shell-decomposition}), we define here {\em the triple-scale transfer} as
\begin{align}
{\mathcal S}(\si|\sii|\siii)  =\bigg{\langle} \frac{T}{2F}(\vv^{[\siii]} \cdot \nabla h^{[\sii]}) h^{[\si]} \bigg{\rangle}\;,\label{triple-transfer}
\end{align}
which measures the energetic interaction between three waveband prescribed scales. As the role of the velocity $\vv$ is to advect the spatial gradients of the distribution $h$, the scales on position $\siii$ have the role of mediating the transfers between scales $\sii$ and $\si$. We furthermore consider that the scale $\si$ receives energy if the transfer is positive, a choice consistent with the interpretation used in past studies~\cite{Teaca:2012p1415, Teaca:2014p1571,Told:2015p1712}. 

Conceptually, considering wavebands of infinitesimal thickness in the continuous limit of the spectral space, we can obtain ${\mathcal S}(k|p|q)$ as the transfer between three wave-norm scales. While we will use this to provide some formal definitions that will be simpler to grasp, {we emphasise that we only have access to the power law waveband decomposition, which is the most efficient choice from a computational point of view.}

\subsection{Properties of the triple-scale transfer}

Assigning specific values to $\si$, $\sii$ and $\siii$, we list the properties for the triple-scale transfer. For each mediator $\siii$, the amount of energy received by scale $\si$ is opposite the amount of energy given by $\sii$,
\begin{align}
{\mathcal S}(\si|\sii|\siii)  =-{\mathcal S}(\sii|\si|\siii)\;. \label{triple-transfer-anti}
\end{align}
This can be shown from eq. (\ref{triple-transfer}), using derivation by parts, accounting for the periodic boundaries employed here and considering that $\nabla \cdot \vv^{[\siii]} \equiv0$, which results from the definition of $\vv$. From the relation (\ref{triple-transfer-anti}) we trivially find that ${\mathcal S}(\si|\si|\siii)=0$, i.e. the energy transferred from one scale to itself is zero. The conservation of energy implies that the sum of all transfers occurring between the same three scales is zero,
\begin{align}
{\mathcal S}(\si|\sii|\siii) + {\mathcal S}(\sii|\si|\siii) + {\mathcal S}(\si|\siii|\sii) + {\mathcal S}(\siii|\si|\sii) + {\mathcal S}(\sii|\siii|\si) + {\mathcal S}(\siii|\sii|\si) =0\; \label{conserv}
\end{align}
and is shown to be true by employing the antisymmetry property given by eq. (\ref{triple-transfer-anti}). All the properties listed above will be inherited by subsequent diagnostics that are constructed on the triple-scale transfers ${\mathcal S}(\si|\sii|\siii)$.

\subsection{The definition of energy transfer diagnostics and the link between them}

For GK turbulence, the {\em scale-to-scale} (shell-to-shell \cite{Verma:2004p206}) transfers have been studied before in the literature~\cite{Tatsuno:2010p1363, BanonNavarro:2011p1274}. They represent one of the first {types} of nonlinear diagnostics to be adopted by the field of plasma turbulence~\cite{Camargo:1995p1564} from the field of hydrodynamical (classical) turbulence~\cite{Domaradzki:1990p145}. {The scale-to-scale transfers
are defined from the triple-scale transfer or directly from a waveband decomposition of the nonlinear term, as  
\begin{align}
{\mathcal P}(\si|\sii) =\sum_{\siii}{\mathcal S}(\si|\sii|\siii) {= \bigg{\langle} \frac{T}{2F}(\vv \cdot \nabla h^{[\sii]}) h^{[\si]} \bigg{\rangle}}\;.
\label{s2s_def}
\end{align}
A scale-to-scale transfer} has the interpretation of the energy received by a scale $\si$ from the scale $\sii$, accounting for all possible mediations. Due to the conservation of {energy}, $\mathcal P(\si|\sii) =-\mathcal P(\sii|\si) $ and $\mathcal P(\si|\si) =0$ for each species. {If $\mathcal P(\si|\sii)$ is determined directly from the nonlinear term, only $N$ calculations (i.e. the number of bands) of the nonlinear term are required. By comparison, ${\mathcal S}(\si|\sii|\siii)$ requires $N^2$ computations of the same type.}

From the scale-to-scale transfers or directly from the triple-scale transfers, we recover the nonlinear {\em transfer {spectrum}}, defined {here} as
 \begin{align}
{\mathcal T}(\si) =\sum_{\sii}{\mathcal P}(\si|\sii) =\sum_{\sii}\sum_{\siii}{\mathcal S}(\si|\sii|\siii)\;.
\label{T_def}
\end{align}
While we can recover the {transfer {spectrum}} from more complex scale decompositions, we mention that knowing the full nonlinear term, required for the incremental integration of the GK equations, is sufficient for the computation of ${\mathcal T}$.

Succinctly, the links between the nonlinear transfer {spectrum}, the scale-to-scale transfers, the triple-scale transfers and the global conservation of free energy by the nonlinear interactions (the sole property needed for their definition), can be summarised as  
\begin{align}
\sum_{\siii}{\mathcal S}(\si|\sii|\siii)&={\mathcal P}(\si|\sii) \;,\\
\sum_{\sii}\sum_{\siii}{\mathcal S}(\si|\sii|\siii)&=\sum_{\sii}{\mathcal P}(\si|\sii) = {\mathcal T}(\si) \;,\\
\sum_{\si}\sum_{\sii}\sum_{\siii}{\mathcal S}(\si|\sii|\siii)&=\sum_{\si}\sum_{\sii}{\mathcal P}(\si|\sii) =\sum_{\si} {\mathcal T}(\si)=0\;.
\label{links_def}
\end{align}
We will present next the {transfer {spectrum}} and the {the scale-to-scale transfers} at a given {instant} in time for the GK simulation analysed in this paper (section \ref{sec:GK:data}).

\subsection{Transfer {spectrum}}

We start by mentioning that {for each species $\sigma$} we approximate the norm $\varepsilon$ in terms of the waveband representation for the transfer spectra, 
\begin{align}
\varepsilon\approx\frac{1}{2}\sum_{\si=1}^{N} \big{|} \mathcal T(\si) \big{|} \;. \label{norm}
\end{align}
This is a practical approximation and the use of the discrete version given by eq. (\ref{norm}) can be seen as being acceptable as long as $\mathcal T(k)$ is not highly fluctuating in a $\si$ interval. 

\begin{figure}[h]
\center
\includegraphics[width = 0.8\textwidth]{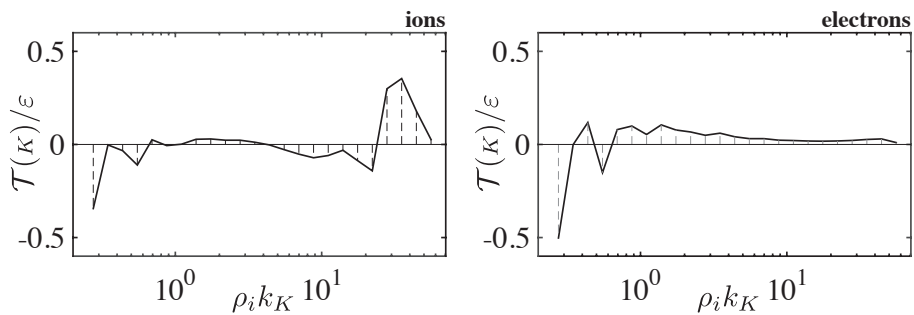}
\caption{The transfer spectra for (left) ions and (right) electrons normalised to their respective $\varepsilon$ values. The vertical dashed lines denote waveband boundaries.}
\label{fig_transfer-spectra}
\end{figure}

For our GK simulation (section \ref{sec:GK:data}) we have $\varepsilon_i=9.65\times 10^1\ [\mbox{GENE} \ power\ units]$ and $\varepsilon_e=2.16\times 10^2\ [\mbox{GENE}\ power\ units]$. Looking at the ratio $\varepsilon_e/\varepsilon_i\approx 2.24$, we see that the electron's energy transfers are {the most intense} in the GK system analysed. {Here, from the transfer spectrum $\mathcal T(\si)$ for ions and electrons (presented in figure~\ref{fig_transfer-spectra}) we observe that the electrons remove more energy from the large (forced) scales. In the absence of time averages, it is hard to properly distinguish properties of the transfer spectra.} 

\subsection{Scale-to-scale transfers}

The scale-to-scale diagnostic provides a way to visualise the energy cascade. Since the waveband boundaries are taken as a power law, the scale-to-scale transfers normalised to their maximal absolute value provide us with information regarding the direction and locality of the energy cascade. We designate a transfer to be direct if it's positive for $\si>\sii$ and we call it local if it occurs primary between scales with $\sii \sim \si$. In section \ref{sec:measurements}, {we will elaborate on the local character of the cascade}.
 
\begin{figure}[t]
\center
\includegraphics[width = 0.8\textwidth]{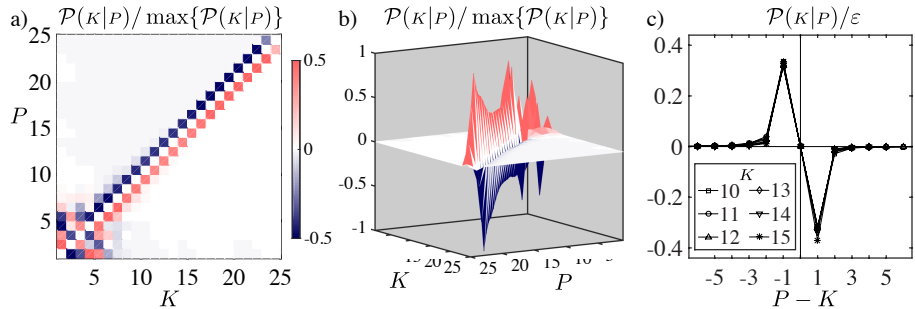}
\caption{Scale-to-scale transfers for the ions: a) flat visualisation; b) pseudo-isometric visualization; c) the collapsed of select $\mathcal P(\si|\sii)$ curves as a function of $\sii-\si$. }
\label{fig_s2s_ions}
\end{figure}

\begin{figure}[t]
\center
\includegraphics[width = 0.8\textwidth]{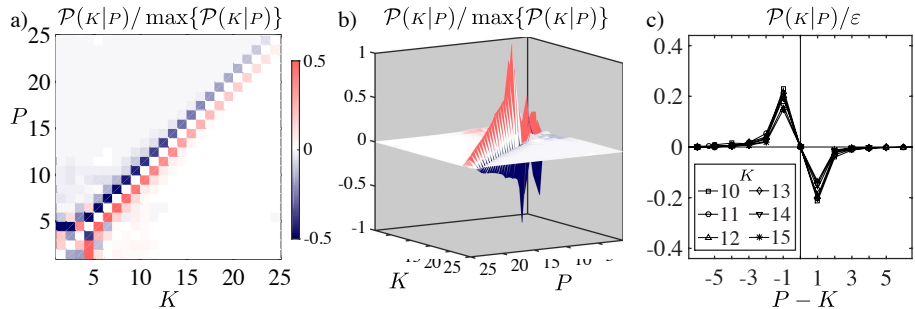}
\caption{Scale-to-scale for the electrons: a) flat visualisation; b) pseudo-isometric visualization; c) the collapsed of select $\mathcal P(\si|\sii)$ curves as a function of $\sii-\si$.}
\label{fig_s2s_electrons}
\end{figure}

From figure~\ref{fig_s2s_ions} for the ions and figure~\ref{fig_s2s_electrons} for the electrons, we do observe that the patterns for the scale-to-scale transfers correspond indeed to a direct and local energy cascade. Since $\mathcal P(\si|\sii)$ is systematically positive for the energy received from larger scales $\si>\sii$ (lower-diagonal {in panels a}), we can say that we observe a direct energy cascade. For both sets of figures, the pseudo-isometric visualisation {(panels b)} allows us to better compare the intensity of the transfers at different scales, while plotting the scale-to-scale transfers as a function of $\sii-\si$ {(panels c)} allows us to better gauge their locality and self-similarity (i.e. the curves for different $\si$'s would collapse on each other for perfect self-similar transfers; an expected property for scale-to-scale transfers in the inertial range of classical homogenous turbulence).  

We {again} mention that it's important to differentiate between the locality of the energy cascade, one structure giving energy to a similar size structure, and the locality of interactions captured by Kraichnan's locality functions \cite{Kraichnan:1965p932}, where the mediators of the energetic interaction between two scales are also considered. While the two are related, as we will show in section \ref{sec:measurements} and discuss in the last section, they are not directly equivalent.

{For the system analysed here ($\beta=1$), we notice that the ion's scale-to-scale transfers tend to be self-similar in a range starting with the gyroradius ($k\rho_i\sim1$) and ending around $k\rho_i\sim20$. The latter limit is due to the collisional dissipation employed, which for the ions is dominated by the field perpendicular contributions (including the $k_\perp$ Finite Larmor Radius - FLR effects)~\cite{Navarro:2016p1965}. Considering the overall shape of $\mathcal T_i$ (large scale source, small scale sink), the ions match the classical picture of turbulence to a certain degree. By comparison, electrons suffer from strong parallel mixing effects at scales $k\rho_i>1$, which help to remove free energy through parallel collisions. For electrons dominated by Landau damping, the role of the perpendicular cascade is just to mop up the reminder of the free energy and pass it down to ever smaller scales, which are in turn also affected by Landau damping. This leads to the "damped" cascade picture observed in figure~\ref{fig_s2s_electrons}. As the electron dissipative route is more efficient than the ion one, there should be no surprise that in a steady state the electron free energy channel draws in more free energy from a given source (leading to $\varepsilon_e/\varepsilon_i\approx 2.24$). }

\subsection{The energy flux across a scale}

Compared to the transfers between different wavebands, the locality functions and scale fluxes are {functions of the waveband boundaries}. Numerically, we have access to the triple-scale transfer $\mathcal S(\si|\sii|\siii)$ and thus, we can compute with ease the scale flux through the scale boundaries ($k_c=k_1 \lambda^c$). In terms of the triple-scale transfer information, the {\it scale flux} reads
\begin{align}
\Pi(k_c) &=\sum_{\si=c+1}^{N} \sum_{\sii=1}^{N} \sum_{\siii=1}^{N} \mathcal S(\si|\sii|\siii) 
=\sum_{\si=c+1}^{N} \sum_{\sii=1}^{N}  \mathcal P(\si|\sii) 
=\sum_{\si=c+1}^{N} \mathcal T(\si) \label{fluxdeff}\ , 
\end{align}
where the last two identities relate the scale flux to the scale-to-scale transfer and the transfer spectrum, respectively. 

\begin{figure}[b]
\center
\includegraphics[width = 0.8\textwidth]{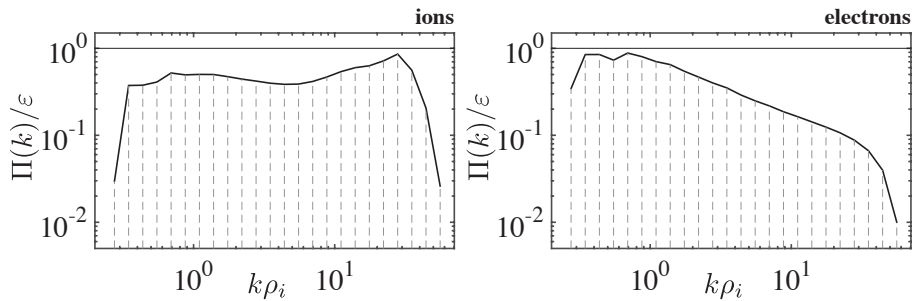}
\caption{The fluxes of energy across a scale for (left) ions and (right) electrons normalised to their respective $\varepsilon$ values. The vertical dash lines denote waveband boundaries.}
\label{fig_flux}
\end{figure}

We display in figure~\ref{fig_flux} the scale flux for the ions and electrons for our GK simulation. While in the context of reduced ($z$ invariant) GK turbulence \cite{Cerri:2014p1757}, the departure from scale invariance for an energy flux was shown to be dependent on the perpendicular collisions, the scaling displayed by the electron flux should be analysed from the perspective of Landau damping, as electrons dissipate energy mainly due to parallel velocity collisions and a $v_\parallel, z$ coupling is established via linear phase mixing. This acts as a reminder that phase space dynamics cannot be ignored, even though such an analysis goes beyond the scope of this paper and {is} left for future work. 

Since definitions listed for a continuous system can {sometimes offer a clearer} understanding, we provide an equivalent definition for the scale flux across the waveband boundaries $k_c$ in terms of the infinitesimal triple-scale transfer $\mathcal S(k|p|q)$,  
\begin{align}
\Pi(k_c) &=  \int_{k_c}^{\infty} \! dk \int_{0}^{\infty} \!  \int_{0}^{\infty} \! dp\, dq\ \mathcal S(k|p|q)=-  \int^{k_c}_{0} \!dk \int_{0}^{\infty}\!  dp \int_{0}^{\infty} \!  dq\ \mathcal S(k|p|q) \nonumber \\
&=  \int^{k_c}_{0}\! dp \int_{0}^{\infty}\! dk  \int_{0}^{\infty} \!  dq\ \mathcal S(k|p|q)
 \label{fluxdefftwo}\ , 
\end{align}
where we have used $\mathcal S(k|p|q)=-\mathcal S(p|k|q)$ in the last equality and remembered that due to eq.~(\ref{conserv}) we have 
\begin{align}
\int_{a}^{b} \! dk \int_{a}^{b}\! dp \int_{a}^{b}\! dq \ \mathcal S(k|p|q)=0\; , \label{intzero}
\end{align}
for any $a$ and $b$ sub-domain limits. Unlike energy transfers that depend on the thickness of the bands, the value of $\Pi(k_c)$ is identical regardless of it being computed via eq.~(\ref{fluxdeff}) or from eq.~(\ref{fluxdefftwo}).

\subsection{The locality functions} \label{sec:loc}

Knowing the scale flux through $k_c$, the infrared (IR) locality function is defined by taking a probe wavenumber boundary $k_\kp$, so that for $k_\kp \le k_\kc$ we have 
\begin{align}
\Pi_{\mbox{\scriptsize ir}}\,\! (k_\kp|k_\kc) \!=\! \! \sum_{\si=\kc+1}^{N} \! \left[\sum_{\sii=1}^{N} \sum_{\siii=1}^\kp \!+\! \sum_{\sii=1}^{\kp} \sum_{\siii=\kp+1}^N  \right] \! \mathcal S({\si|\sii|\siii})\;. \label{IRloc}
\end{align}
The IR locality function measures the contribution to the flux through $k_\kc$ from couplings with at least one scale wavenumber less than $k_\kp$. In the second term, the sum over waveband $\siii$ starts from $\kp+1$ to avoid double counting. In the limit $k_\kp \rightarrow k_\kc$, we recover the flux across the cutoff wavenumber $k_c$, {i.e.
\begin{align}
\Pi_{\mbox{\scriptsize ir}}\,\! (k_\kp|k_\kc) \!&=\! \! \sum_{\si=\kc+1}^{N} \! \left[\sum_{\sii=1}^{N} \sum_{\siii=1}^\kp \mathcal \!+\! \sum_{\sii=1}^{\kp} \sum_{\siii=\kp+1}^N  \right] \! \mathcal S({\si|\sii|\siii})\; \nonumber\\
&=\! \! \sum_{\si=\kc+1}^{N} \! \left[\sum_{\sii=1}^{N}  \bigg{(} \sum_{\siii=1}^N \!-\!\! \sum_{\siii=\kp+1}^N
\bigg{)} \!+\! \bigg{(} \sum_{\sii=1}^{N} \!-\!\! \sum_{\sii=\kp+1}^{N} \bigg{)} \sum_{\siii=\kp+1}^N  \right] \! \mathcal S({\si|\sii|\siii})\; \nonumber \\
&=\! \! \sum_{\si=\kc+1}^{N} \! \sum_{\sii=1}^{N}  \sum_{\siii=1}^N \! \mathcal S({\si|\sii|\siii}) -\!\!\sum_{\si=\kc+1}^{N}  \sum_{\sii=\kp+1}^{N} \sum_{\siii=\kp+1}^N  \! \mathcal S({\si|\sii|\siii}) \; \label{IRmanip}
\end{align}
recovers eq.~(\ref{fluxdeff}) for $p\!=\!c$ due to the conservation of energy 
\begin{align}
\sum_{\si=a}^{b}  \sum_{\sii=a}^{b} \sum_{\siii=a}^b   \mathcal S({\si|\sii|\siii})=0\, ,
\end{align}
for any $a, b$ indices that identify a set of bands.} It is customary to normalise the locality functions to the flux trough $k_c$, in which case a value of one is obtained for $k_p=k_c$ and less than one for $k_p/k_c<1$. 

Although the IR functions have a clear interpretation as the ratio of energy contributed to the flux through scale $k_\kc$ coming only from larger and larger scales, it should be remembered that for $k_p/k_c\ll1$ the transfers can only take place between triads with one wave vector leg much smaller compared to the other two. Therefore, these functions can provide information regarding the overall locality of the nonlinear interaction. {The rate with which $\Pi_{\mbox{\scriptsize ir}}\,\! (k_\kp|k_\kc)/\Pi\! (k_\kc)$ decreases in value as a function of $k_p/k_c$ measures the locality of interactions. Larger exponents denote more local behavior while a zero exponent would qualify turbulence as fully nonlocal, i.e. every scale influencing equally the coupling of any other scale in the system.}

\begin{figure}[t]
\center
\includegraphics[width = 0.8\textwidth]{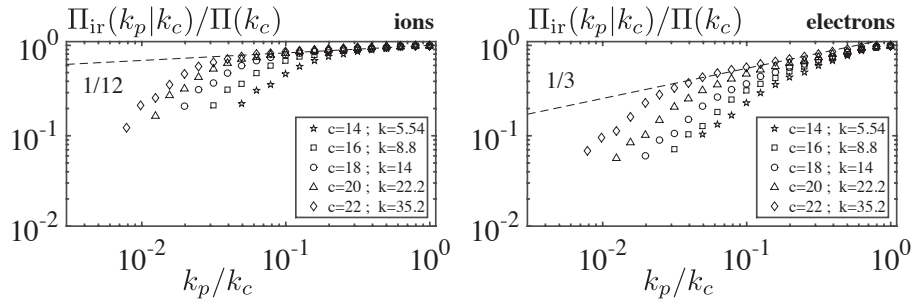}
\caption{The IR locality functions for (left) ions and (right) electrons normalised to the energy flux.}
\label{figIR}
\end{figure}

\begin{figure}[b]
\center
\includegraphics[width = 0.8\textwidth]{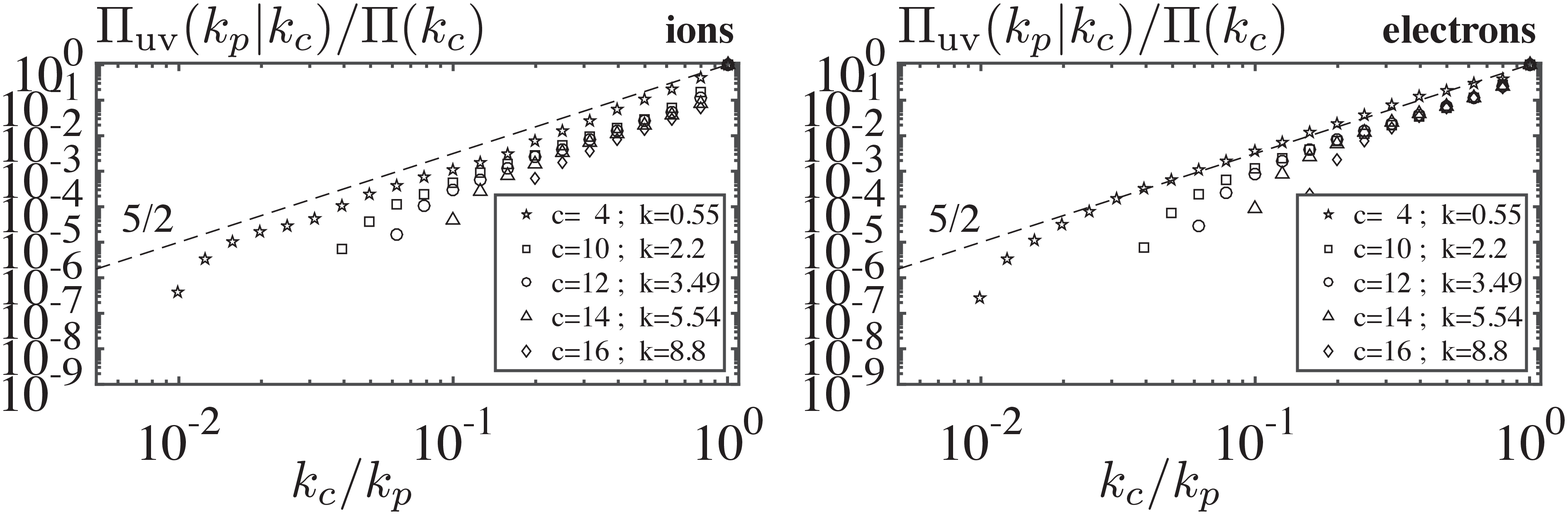}
\caption{The UV locality functions for (left) ions and (right) electrons normalised to the energy flux.}
\label{figUV}
\end{figure}

For our simulation data, the IR locality is presented in figure~\ref{figIR}. The $1/12$ value for the ion scaling exponent is consistent with the value reported by Ref.~\cite{Teaca:2012p1415} in the first study of the locality {problem} for GK turbulence, which was performed for tokamak-relevant magnetic geometry. The $1/3$ value for the electron's IR scaling was already shown recently by Ref.~\cite{Told:2015p1712}, on the same data. We mention that the exponents' values appear to be much smaller (more nonlocal behavior) than for the hydrodynamic~\cite{Zhou:1993p949, Eyink:2005p854} (i.e. $4/3$) or the MHD~\cite{Aluie:2010p946,Domaradzki:2010p1012, Teaca:2011p1362} (i.e. $2/3$) cases. 

{Using as reference the $4/3$ value found for the asymptotic locality exponent in classical turbulence~\cite{Zhou:1993p949, Eyink:2005p854}, we will associate larger values for the GK exponents to {\it local} and smaller values to {\it nonlocal} interactions. While this is an arbitrary choice, this simple characterisation will allow us to simplify the presentation of the results.}

A similar definition is made for the ultraviolet (UV) locality functions, which for $k_\kc \le k_\kp$ is given as 
\begin{align}
\Pi_{\mbox{\scriptsize uv}}\,\! (k_\kp|k_\kc) \!=\! \! \sum_{\sii=1}^{\kc} \! \left[\sum_{\si=\kp+1}^{N} \sum_{\siii=1}^N \!+\! \sum_{\si=1}^{\kp} \sum_{\siii=\kp+1}^N \right] \! \mathcal S({\si|\sii|\siii} )\;.
\end{align}
It measures the contribution to the flux through $k_\kc$ from couplings of scales with at least one scale wavenumber greater than $k_\kp$, therefore providing information regarding the locality makeup of a scale $k_\kc$ in relation with smaller and smaller scales. For our simulation the UV locality is presented in figure~\ref{figUV}. A strong UV locality character is inferred from the large values of the UV locality exponents. We mention that selecting a cutoff {$k_c\rho_i=0.55$} gives the same scaling as for {$k_c\rho_i>1$}, denoting that gyro-averaging effects don't influence the contributions to the flux across a scale emerging from interactions with progressively smaller scales. For both plasma species, a robust $5/2$ value can be inferred for the UV locality exponent. 

For completeness, as we will use them in later sections, we provide the definitions of the locality functions in terms of the infinitesimal triple-scale transfer $\mathcal S(k|p|q)$,  
\begin{align}
\Pi_{\mbox{\scriptsize ir}}\,\! (k_\kp|k_\kc)  &=  \int_{k_c}^{\infty} dk \left[\int_{0}^{k_\kp}  dp \int_{p}^{\infty}   dq\ \mathcal S(k|p|q) + \int_{0}^{k_\kp}  dq \int_{q}^{\infty}   dp\ \mathcal S(k|p|q)\right]  \label{IRtwo}\ , \\
\Pi_{\mbox{\scriptsize uv}}\,\! (k_\kp|k_\kc) &=  \int_{0}^{k_c} dp \left[\int_{k_p}^{\infty}  dk \int^{\infty}_{0}   dq\ \mathcal S(k|p|q) + \int_{0}^{k_p}   dk  \int^{\infty}_{k_\kp}  dq\ \mathcal S(k|p|q)\right]  \label{UVtwo}\ . 
\end{align}
{These definitions are analogous to their waveband discrete forms, up to a manipulation of the IR integral limits similar to the one performed in eq.~(\ref{IRmanip}) and by employing eq.~(\ref{intzero}).}

\subsection{Reviewing the significance of the results}

The results listed for the transfers, fluxes and locality functions have been, in one form or another, presented in the literature. We take the time to comment on their significance, before performing a more detailed analysis on the transfers. 

The fact that the energy transfers are dominant between neighboring wavebands denote the local character of the  cascade. {However, considering that we take wavebands separated by less than a factor $2$ (i.e. $2^{1/3}$), the observed energy exchanges imply strong intra-dyadic interactions, known to lead to an enhanced nonlocal characteristic for the nonlinear interactions} (we will detail this statement in section \ref{sec:measurements}). {This assertion is validated by the strong nonlocal character captured by the IR locality functions (i.e. exponents reaching values of $1/12$ for the ions and $1/3$ for the elections, values much smaller than the $2/3$ and $4/3$ found in MHD and hydrodynamical turbulence).} The locality nature of interactions is important when modeling turbulence \cite{BanonNavarro:2014p1535}, but it also has a strong phenomenological significance. A strong IR non-locality is a sign of a very large scale flow shearing small scale structures, rather than advecting them. {For example, a zonal flow~\cite{Diamond:2005p222} is expected to enhance the non-locality nature of plasma turbulence.} On the other hand, the local UV behavior can be seen as an insensitivity of the perpendicular {interactions} to the type of collisional operators employed. {This is fortunate from the perspective of turbulence modeling, even more so when we consider that both ions and electrons recover the same robust $5/2$ value for the UV locality exponent, although their phase space dissipation route is found to be different \cite{Navarro:2016p1965}. }

{In the case of IR locality functions (figure~\ref{figIR}) we also observe a change in slopes at $k_p\rho_i\sim1$, an effect we consider specific to magnetised plasma turbulence, as turbulence above and below the ion gyroradius is expected to have different properties.}
The IR exponent seems to be much larger (more local behavior) at scales $k\rho_i<1$. The ions strong nonlocal behavior {occurs at scales} $k\rho_i>1$. As IR non-locality increases, the transfers associated with intra-dyadic interactions become more important in the energy cascade. As these intra-dyadic transfers, {in the limit $q\rightarrow 0$,} can be seen as occurring between neighboring waves rather than {compact structures}, the GK turbulence cascade may {have a stronger wave characteristic}~\cite{Howes:2011p1459} compared to strong classical turbulence {or even MHD turbulence}.

{For the remainder of this paper, we seek to understand better the link between dyadic separated couplings and intra-dyadic interactions and their contribution to the locality problem.}

\section{Idealised energy transfers between scales}  \label{sec:test}

While diagnostics based on $\mathcal S(\si|\sii|\siii)$ (e.g. the locality functions) can offer the most amount of information pertinent to the nonlinear interactions, the triple-scale transfers can be very demanding to compute, especially for the five-dimensional gyrokinetic problem. For a scale decomposition that uses {$N$ wavebands}, the triple-scale transfers require $N^2$ calculations of the nonlinear terms. By comparison, computing {the scale-to-scale transfers $\mathcal P(\si|\sii)$ directly from the nonlinear term requires only $N$ nonlinear terms calculations}. Since $\mathcal P(\si|\sii)$ measures the energy exchange between two scales, a natural question emerges: can we recover information pertinent to scale locality directly from $\mathcal P(\si|\sii)$? And if yes, what is its interpretation? 

\subsection{Definition}

To address these questions and to build confidence in the interpretation of diagnostics that {are applied} in practice to turbulence systems that don't exhibit clear inertial ranges, we define a test case that will prove helpful in these matters. We choose:  
\begin{eqnarray}
\mathcal P^{\mbox{\scriptsize{test}}}(\si|\sii) =\left\{ \begin{array}{lcl}
\ \ \, \Big{(}\frac{k_P}{k_K}\Big{)}^{{\alpha}_{\mbox{\scriptsize ir}}} \, , &  k_P \le k_K  \\
\\
-\Big{(}\frac{k_K}{k_P}\Big{)}^{{\alpha}_{\mbox{\scriptsize uv}}}, &  k_P > k_K  
\end{array}  \right.\;,
\end{eqnarray}
where ${{\alpha}_{\mbox{\scriptsize ir}}}$ and ${{\alpha}_{\mbox{\scriptsize uv}}}$ are here the two control parameters, together with the implicit choice for the wavebands. As the indices' suffixes imply, we take the ${{\alpha}_{\mbox{\scriptsize ir}}}$ and ${{\alpha}_{\mbox{\scriptsize uv}}}$ exponents to be related to the infrared and ultraviolet locality exponents.

\begin{figure}[b]
\center
\includegraphics[width = 0.8\textwidth]{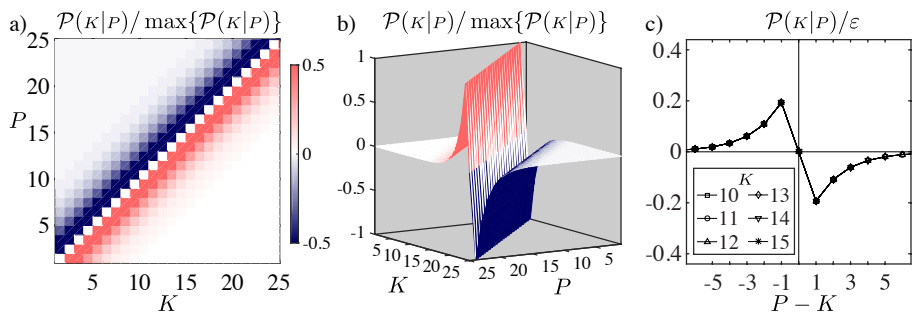}
\caption{Scale-to-scale for the test function ($\mathcal P\equiv\mathcal P^{\mbox{\scriptsize{test}}}$): a) flat visualization; b) pseudo-isometric visualisation; c) the collapsed of select $\mathcal P(\si|\sii)$ curves as a function of $\sii-\si$. Here $\varepsilon=\frac{1}{2}\sum_\si\big{|}\sum_\sii \mathcal P^{\mbox{\scriptsize{test}}}(\si|\sii) \big{|} $ in accordance with eq.~(\ref{norm}).}
\label{fig_s2s_test}
\end{figure}

\begin{figure}[t]
\center
\includegraphics[width = 0.4\textwidth]{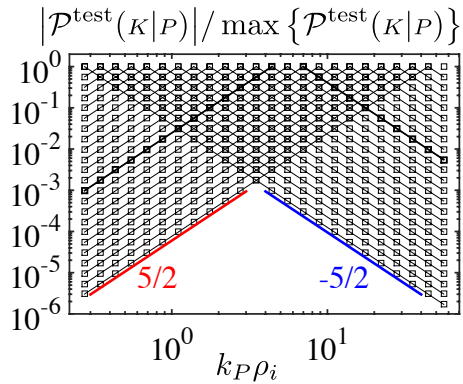}
\caption{$\big{|}\mathcal P^{\mbox{\scriptsize{test}}}(\si,\sii)\big{|}$ for all possible $\si$. We emphasise the transfer for $\si=14$ as a thick line, to help visualise the typical $\si$ transfer curve. The slope of the tails for $\si >\sii$ and $\si <\sii$ are indicated for reference, recovering the $5/2$ value prescribed for ${{\alpha}_{\mbox{\scriptsize ir}}}$ and ${{\alpha}_{\mbox{\scriptsize uv}}}$.}
\label{fig_s2s_test_log}
\end{figure}

Below, we use the same waveband decomposition as the one given in section \ref{sec:scales:waveband} for the GK problem (i.e. $\lambda=2^{1/3}$) and we use ${{\alpha}_{\mbox{\scriptsize ir}}}={{\alpha}_{\mbox{\scriptsize uv}}}=5/2$, to start. The resulting $\mathcal P^{\mbox{\scriptsize{test}}}(\si|\sii)$ transfers are presented in figure~\ref{fig_s2s_test}. Not surprisingly, we recover an idealised forward cascade that is scale-invariant (excepting the start and end of the band interval considered, due to numerical truncation effects). The scale invariance of the transfers is best seen from the figure~\ref{fig_s2s_test}-c) panel, where a perfect collapse of the scale-to-scale transfers as a function of $\sii-\si$ is observed. Moreover, from the same panel we observe that the "tails" of the transfers decreases gradually at $\pm (\sii-\si)$.

In figure~\ref{fig_s2s_test_log} we plot the absolute value of $\mathcal P^{\mbox{\scriptsize{test}}}(\si|\sii)$ as a function of $k_P$ for all $\si$ values. We clearly see that {for this simple test} the tails follow a power law. {We mention that from these type of plots, especially when considering our graphical representation, we only seek to identify an overall power law for the tails, rather than analyse individual transfers characteristics.} The slope of the tails for $\si >\sii$ and $\si <\sii$ are indicated for reference, recovering the values prescribed for ${{\alpha}_{\mbox{\scriptsize ir}}}$ and ${{\alpha}_{\mbox{\scriptsize uv}}}$ parameters. While this should not come as a surprise due to our choice for $\mathcal P^{\mbox{\scriptsize{test}}}(\si|\sii)$, we want to recover these exponents by means of locality functions.

\subsection{Capturing the locality of the test transfers via locality type functions}

\begin{figure}[b]
\center
\includegraphics[width = 0.5\textwidth]{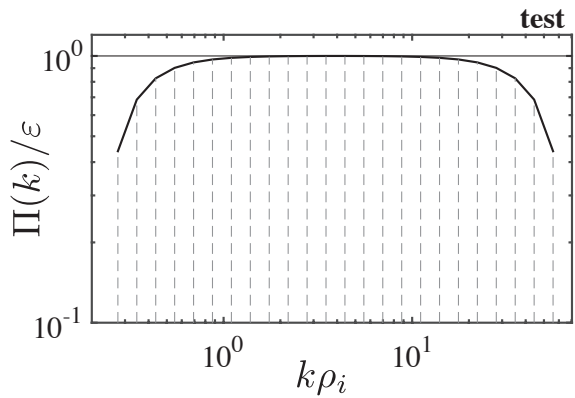}
\caption{The flux of energy across a scale for  the test transfers. The fall-off  at low and high $k$ are due to the finite size effects not being able to account for the locality.}
\label{fig_flux_test}
\end{figure}

\begin{figure}[b]
\center
\includegraphics[width = 0.8\textwidth]{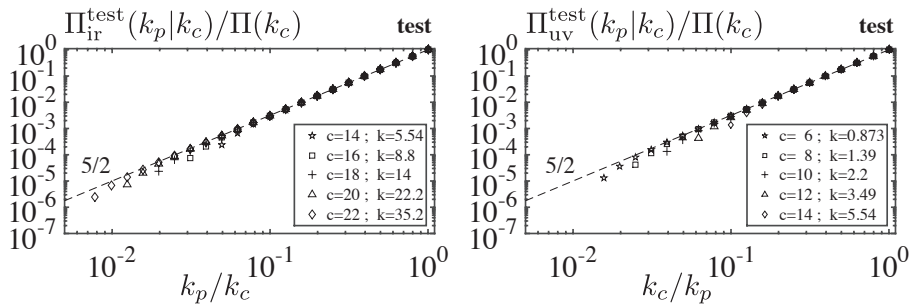}
\caption{The modified IR and UV locality functions for the test transfers ${{\alpha}_{\mbox{\scriptsize ir}}}={{\alpha}_{\mbox{\scriptsize uv}}}=5/2$.}
\label{fig_IR_UV_test}
\end{figure}

\begin{figure}[t]
\center
\includegraphics[width = 0.8\textwidth]{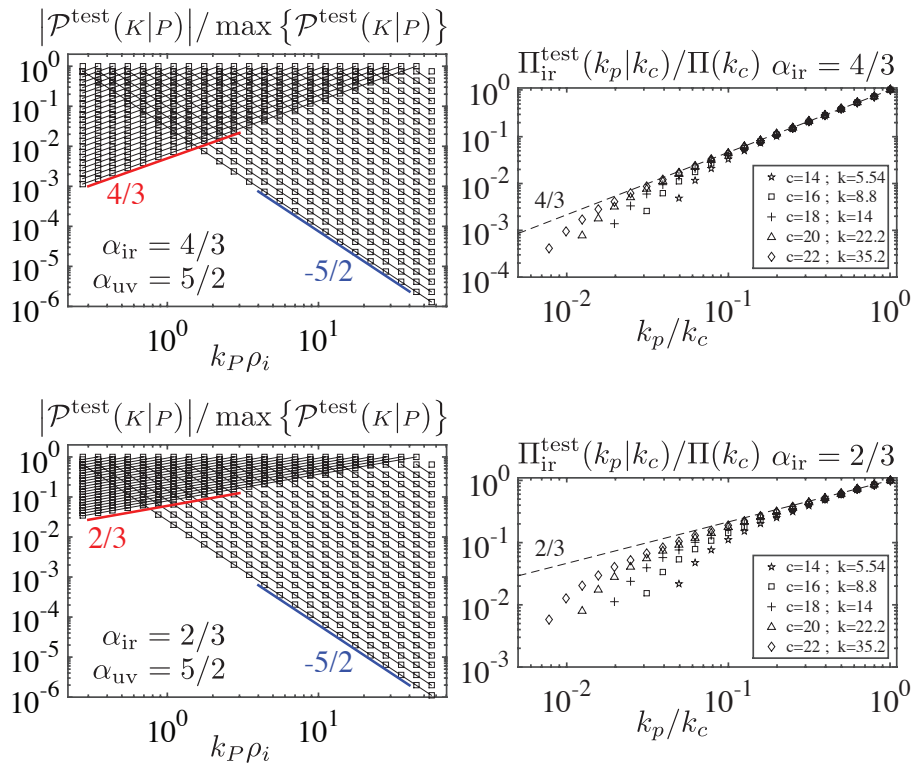}
\caption{The $\big{|}\mathcal P^{\mbox{\scriptsize{test}}}(\si,\sii)\big{|}$ transfers and the modified IR locality functions for differnt values of ${{\alpha}_{\mbox{\scriptsize ir}}}$. For all cases, the UV locality functions are identical as the one given in Figure~(\ref{fig_IR_UV_test})}.
\label{fig_locality_test}
\end{figure}

Taking into account that we can recover from $P^{\mbox{\scriptsize{test}}}(\si|\sii)$ the flux through a surface $k_c$ (displayed in figure~\ref{fig_flux_test}) by simply employing the definition given by eq.~(\ref{fluxdeff}), i.e.
\begin{align}
\Pi^{\mbox{\scriptsize{test}}}(k_c)=\sum_{\sii=1}^{c}\sum_{\si=c+1}^{N} \mathcal P^{\mbox{\scriptsize{test}}}(\si|\sii) \;  \label{fluxtest}
\end{align}
{where $\sum_{\sii=c+1}^{N}\sum_{\si=c+1}^{N} \mathcal P^{\mbox{\scriptsize{test}}}(\si|\sii)=0$ due to energy conservation},  we take an additional probe surface that limits the separation between the energy giving and the energy receiving scales. In the absence of a mediator, we cannot recover the IR and UV locality functions previously introduced and use the following definitions instead: 

\begin{align}
& \Pi^{\mbox{\scriptsize{test}}}_{\mbox{\scriptsize ir}}(k_p | k_c)=\sum_{\sii=1}^{p}\sum_{\si=c+1}^{N} \mathcal P^{\mbox{\scriptsize{test}}}(\si|\sii)\; \ , \mbox{for} \  k_p\ge k_c, \\
& \Pi^{\mbox{\scriptsize{test}}}_{\mbox{\scriptsize uv}}(k_p | k_c)=\sum_{\sii=p+1}^{N}\sum_{\si=1}^{c} \mathcal P^{\mbox{\scriptsize{test}}}(\si|\sii)\; \ ,  \mbox{for} \  k_p\le k_c. 
\end{align}
We see that in the limit $k_p \rightarrow k_c$, the two locality functions recover the energy flux given by eq.~(\ref{fluxtest}). The test locality function captures the prescribed exponents, figure~\ref{fig_IR_UV_test}. This is also seen from the two additional cases presented in figure~\ref{fig_locality_test}, where we use the same value ${{\alpha}_{\mbox{\scriptsize uv}}}=5/2$ but different values for ${{\alpha}_{\mbox{\scriptsize ir}}}$. We see that the values prescribed are recovered by the {test} locality exponents. We mention that the UV plots are exactly the same as the one listed in figure~\ref{fig_IR_UV_test}. 

A few lessons can be drawn from this simple test. First, {this simple test allows us to clearly identify characteristics we associate with asymptotic locality, i.e. for different values of $k_c$ all curves collapse on each other and exhibit the same power law. Second, for more nonlocal scalings (small value of the locality exponent), the} locality functions fall off the asymptotic scaling at small $k_p/k_c$ ratios. This fall-off is due to the high non-locality nature not being able to isolate the finite domain fast enough and limits our ability to gauge the correct IR locality behavior close to the largest scale. The slope at high $k_p/k_c$ values is where asymptotic values should be investigated. We mention that for ${{\alpha}_{\mbox{\scriptsize ir}}} \ne {{\alpha}_{\mbox{\scriptsize uv}}} $ the test transfer considered does not sum up to zero. This is a simple particularity of the definition employed, as there is no a-priori requirement for IR and UV locality exponents to be identical. This is acceptable as we just use the different ${{\alpha}_{\mbox{\scriptsize ir}}}$ values to test the ability of the modified locality functions to capture the correct IR exponents. 

The most important lesson to be drawn is also the most obvious: The locality exponents can be determined directly from the scale-to-scale transfer. However, these are not the exponents related to all interactions, but related to the locality of the energy cascade. We go back to the full GK simulation to show this fact.

\section{Detailed measurements of the energetic exchanges in GK turbulence} \label{sec:measurements}

Compared to the idealised test case that was presented in the previous section, when investigating nonlinear energetic interactions and their scale locality in turbulence, we need to account implicitly or explicitly for the contribution made by the mediator scale. As an example, we start by looking at the ion scale-to-scale transfers for various separations between the energy exchanging scales.

\subsection{The impact made by the separation of scales on the ion scale-to-scale transfers}

{The scale-to-scale transfers (eq.~\ref{s2s_def}) are obtained by integrating over all mediator scales. Here, rather than doing so, we will characterise the exchanges between two scales $p$ and $k$ as a function of the values taken by the mediators $q$, and integrate accordingly over $q$ to obtain a conditional form of the scale-to-scale transfers. } 

The minimal value for a scale $q$ that mediates the interaction of two other scales ($p$ and $k$) is given by eq.~(\ref{qmin}), i.e. $q \ge \max\{k,p\}-\min\{k,p\}$. Expressing the value of $\max\{k,p\}$ in terms of $\min\{k,p\}$, as $\max\{k,p\} =\alpha \min\{k,p\}$ with $\alpha \ge 1$, we obtain
\begin{align}
q \ge (\alpha-1) \min\{k,p\} \;.
\label{qlim}
\end{align}
{We see that the lower limit of the mediator $q$ (i.e. $q_{\min}=(\alpha-1) \min\{k,p\}$) and the minimal separation between $p$ and $k$ are linked. For $k=p$, we in fact have $\alpha =1$ and $q_{\min}=0$. Imposing $q\ge q_{\min}=0$ as a selection condition would pick up all possible scale couplings for which $\max\{k,p\} \ge \min\{k,p\}$ (i.e. any $p$ and $k$ would lead to a $q$ which is larger than $q_{\min}$). This includes the $k=p=q$ couplings. As another example, choosing $\max\{k,p\} = 2 \min\{k,p\}$, which means $\alpha =2$, leads to $q_{\min} = \min\{k,p\}$. Imposing the selection condition $q_{\min} = \min\{k,p\}$, selects all scale couplings ($q\ge q_{\min}$) for which $\max\{k,p\} \ge 2\min\{k,p\}$. Imposing the lower limit of the mediator $q$ sets the the minimal separation possible between the energy exchanging scales.}

{We see that in principle, $\alpha$ can act as a control parameter. For $\alpha=1$, which recovers $k=p$, we have $q_{\min} = 0$. Taking $\alpha=2$ leads to $q_{\min}=\min\{k,p\}$ and ensures that a dyadic separation (i.e. $\max\{k,p\}=2\min\{k,p\}$) is the minimal separation available between the energy exchanging scales. Going further with our example, for $\alpha=3$ we obtain $\max\{k,p\}=3\min\{k,p\}$ as the minimal separation between energy exchanging scales and $q_{\min} \ge 2\min\{k,p\}$.}

{Since our scales are prescribed as multiples of $\lambda =2^{1/3}$, e.g. $k_\siii=k_1\lambda^{\siii-1}$, the values that $\alpha$ can take {are discrete and considered here as} $\alpha=\lambda^n+1$ for $n \in \mathbb{Z}$. Considering that the $\siii$ index identifies the waveband $s_\siii=(\lambda^{-1}k_{\siii},\ k_\siii]$, we write the condition given by eq.~(\ref{qlim})} for the waveband boundaries (i.e. $q_\siii \ge \lambda^n \min\{k_\si, p_\sii \}$) in terms of the waveband indices as 
\begin{align}
\siii \ge \min\{\si, \sii\}+1+n \;. \label{Qcond}
\end{align}
{The $+1$ emerges due to our choice for the waveband index (i.e. $(k_{\siii-1},\ k_\siii]$) and the integer $n$ can now be used as the control parameter for our discrete waveband representation. Corresponding to $\alpha=2$ we have $n=0$ and $\siii_{\min} = \min\{\si, \sii\}+1$, while for the previous $\alpha=3$ example we have $n=3$ and $\siii_{\min} = \min\{\si, \sii\}+4$.} Not only that we {can recover a minimal dyadic separation} between energy exchanging scales for $n=0$, {but we can also select a minimal separation that is less than dyadic} for $n<0$. 

\begin{figure}[t]
\center
\includegraphics[width = 0.8\textwidth]{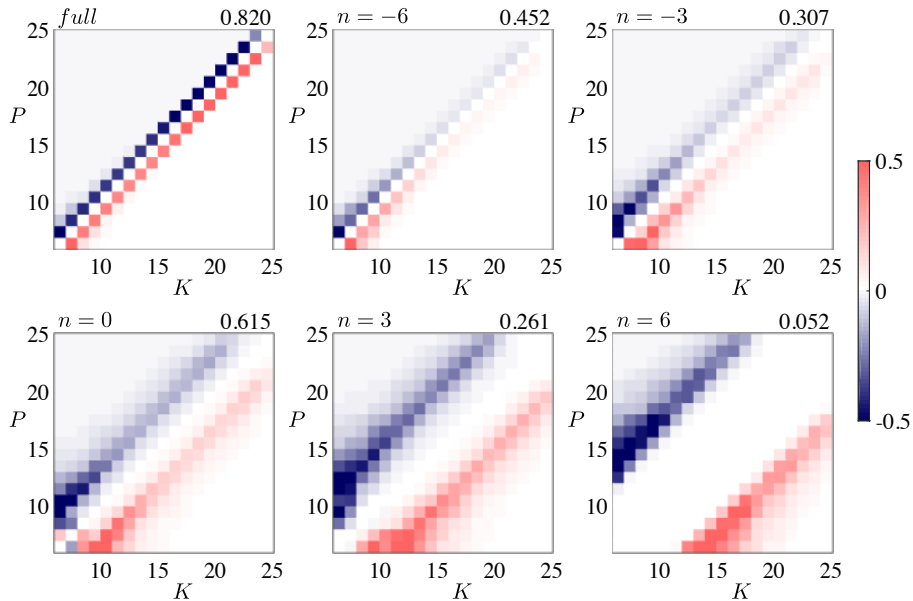}
\caption{The $\widetilde {\mathcal P}^{n}(\si|\sii)$ transfers for the ions, normalised to their respective maximal value, i.e. $\max\{\widetilde {\mathcal P}^{n}(\si|\sii)\}$. The first top panel depicts the standard transfers $ {\mathcal P}(\si|\sii)$, while the rest are identified by the value of $n$. For each panel, the top right number represents the maximal value used for normalisation in units of the transfer norm $\varepsilon$.}
\label{fig_S2SiSeparation}
\end{figure}

Accounting for the condition expressed by eq.~(\ref{Qcond}), we define the modified scale-to-scale transfers as 
\begin{align}
 \widetilde {\mathcal P}^{n}(\si|\sii) =\!\!\!\!\! \sum_{\siii =\min\{\si, \sii\}+1+n}^N \!\!\!\!\!   {\mathcal S}(\si|\sii|\siii)
\label{s2s_separation}\;
\end{align}
and plot them {in figure~\ref{fig_S2SiSeparation} for different values of $n$, using the ion species as an example}. We see that for {a separation between energy exchanging scales that is at least dyadic} ($n=0$) the index separation between the intense peaks and the diagonal is $3$ (since $2 k_\siii=k_{(\siii+3)}$). As expected, selecting larger values for $n$ {limits the selection of energy exchanges between scales that are farther and farther apart.} 

\begin{figure}[t]
\center
\includegraphics[width = 0.8\textwidth]{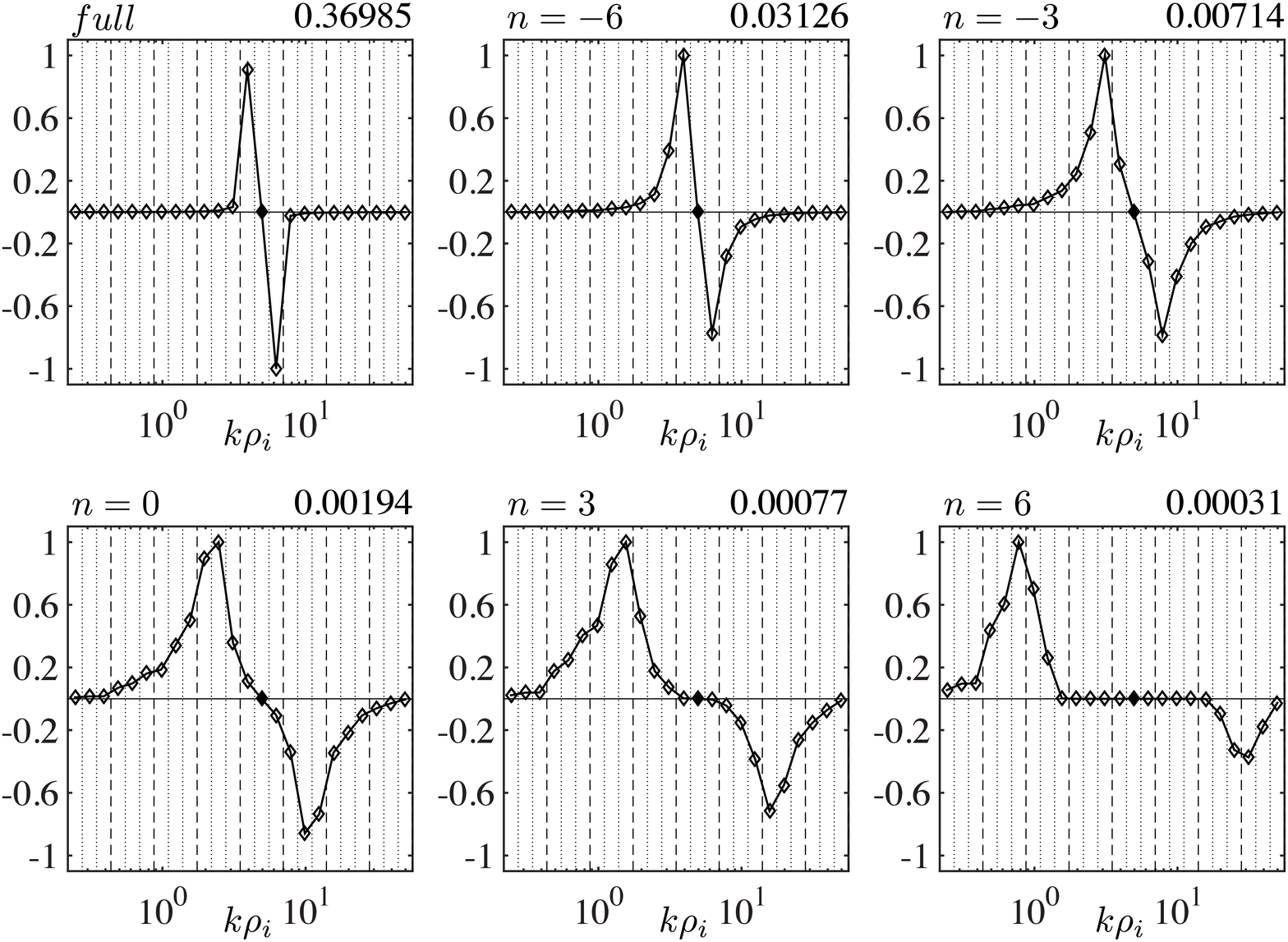}
\caption{The $\widetilde {\mathcal P}^{n}(\si|\sii)$ ion transfer as a function of $k_\sii$ for $\si=14$, normalised to their respective maximal value displayed. The location of $k_1 2^{n/3}$ shell boundaries are depicted by dotted lines and every third boundary, corresponding to a dyadic separation of the form $k_1 2^{n}$, are emphasised as dashed lines. The first top panel depicts the standard transfers $ {\mathcal P}(\si|\sii)$, while the rest are identified by the value of $n$. For each panel, the top right number represents the maximal value used for normalization in units of the transfer norm $\varepsilon$.}
\label{fig_S2SiSide}
\end{figure}

For clarity, we present in figure~\ref{fig_S2SiSide} the same transfers, but only for a selected receiving waveband $\si=14$. We see that only for $n=0$, the scale $k_\si$ is dyadic separated on each side from $k_\sii$, {i.e. from the black diamond we need to count $3$ diamonds (due to $\lambda=2^{1/3}$) to reach the either left or right peaks}. Compared to the full scale-to-scale transfers that are dominated by the {intra-dyadic exchanges}, the {dyadic separated} transfers have a wider $k$ support and decay to zero more gradually. This is important, since the decay of the transfers for $\sii \ll \si$ and $\sii \gg \si$ contains {information relevant to the locality of the cascade}, as we saw for the ideal case. We also note that for a larger separation between the energy exchanging scales, the transfers decrease in intensity, a fact seen from the diminishing value for the ratio between the maximal value of the transfer and the overall transfer norm $\varepsilon$. {The results presented in figures~\ref{fig_S2SiSeparation} and \ref{fig_S2SiSide} validates our expectation that limiting the minimal value of the possible moderators $q$ results in an effective limitation of the minimal separation between the exchanging scales $p$ and $k$.}

{From the $n=-6$ case (for which {$q_{\min} = \frac{1}{4}\min\{k,p\}$ and the minimal separation is} $\max\{k,p\} =\frac{5}{4} \min\{k,p\}$) we can infer an additional important result. We see that we allow for a relatively small separation between the $k$ and $p$ exchanging scales and since $q\in[\frac{1}{4}\min\{k,p\}, \frac{6}{4}\min\{k,p\}]$, the $q\sim k$ or $q\sim p$ values are possible for the mediators. We can consider this as a proxy for the intensity of the energy exchanges for the $k\sim p \sim q$ subset of couplings. By comparison, the {\it full} case (i.e. $q\ge0$), while allowing for these exchanges to take place, it also accounts for the exchanges between $k\sim p$ mediated by $q\ll k\sim p$. In figure~\ref{fig_S2SiSide}, comparing the intensity of the {\it full} transfers with the ones for the $n=-6$ case, we see (from the maximal value used for normalisation in units of the transfer norm $\varepsilon$) that the latter (i.e. $n=-6$) are ten times smaller in value while exhibiting the same overall location for the peaks. This shows that the dominant intra-dyadic exchanges are the ones mediated by small values of $q$ (i.e. $q\ll k\sim p$), which are intrinsically nonlocal from the perspective of the locality of interactions.}

\subsection{{Extracting locality information from the scale-to-scale measurements}}

\begin{figure}[b]
\center
\includegraphics[width = 0.8\textwidth]{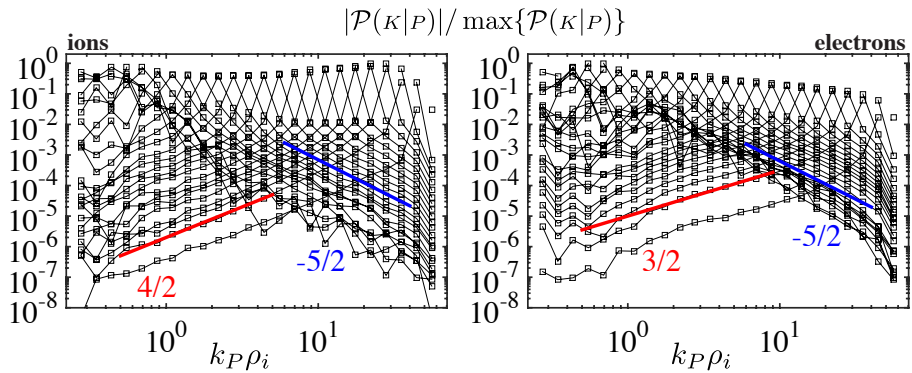}
\caption{$\big{|}\mathcal P(\si,\sii)\big{|}$ for all possible $\si$ for (left) ions and (right) electrons. The slope of the IR for $\si >\sii$ and UV $\si <\sii$ exponents are indicated for reference.}
\label{fig_GKs2s_log}
\end{figure}

\begin{figure}[t]
\center
\includegraphics[width = 0.8\textwidth]{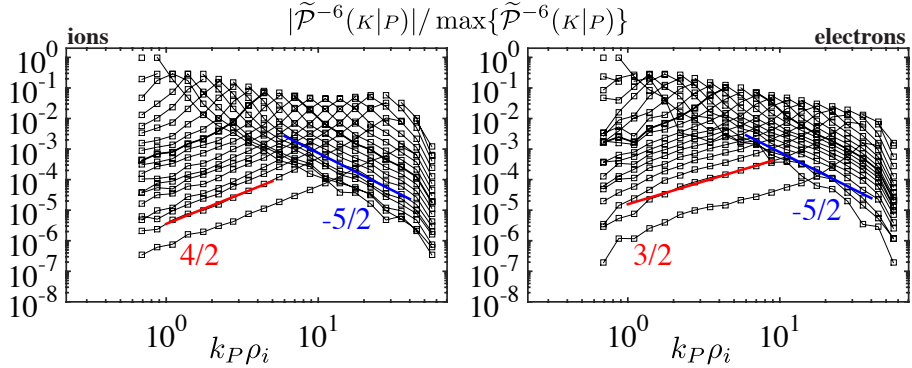}
\caption{$\big{|}\mathcal P(\si,\sii)\big{|}$ for all possible $\si$ for (left) ions and (right) electrons. The slope of the IR for $\si >\sii$ and UV $\si <\sii$ exponents are indicated for reference. We emphasise the transfer for $\si=14$ as a thick line, to help visualise the typical $\si$ transfer curves.}
\label{fig_GKs2s_log2}
\end{figure}

{We seek to extract information pertinent to the locality problem from the scale-to-scale transfers. For ions and electrons,} we look in figure~\ref{fig_GKs2s_log} at the absolute value of $\mathcal P(\si|\sii)$ as a function of $k_P$ for all $\si$ values. While compared to the ideal ({test}) case (section~\ref{sec:test}) we see {that the intra-dyadic} exchanges lead to an abrupt fall-off of the transfer curves (also seen in figure~\ref{fig_S2SiSide}, {\it full} panel) and we see a large and small scale pollution, we are able to identify {(IR and UV)} slopes for the tails. In fact, looking at the $\mathcal P^{-6}(\si|\sii)$ case in figure~\ref{fig_GKs2s_log2}, while noting that the missing points are due to our inability to compute eq.~(\ref{s2s_separation}) for them, we see that the abrupt fall-off is removed while the tail information remains identical. 
{The tail information, or its scaling to be more exact, is what determines the locality information. We also mention that, while $\mathcal P^{-6}(\si|\sii)$ removes the abrupt fall-off, only once the transfers occur between scales more separated (close to a dyadic separation) do we see them fall on a (IR or UV) power law (this can be inferred from the inspection of the thick black lines in figure~\ref{fig_GKs2s_log2}). This is why the dyadic separation represents a useful reference. While not crucial, as the tails can be recovered from the full scale-to-scale transfers, they help identify the correct scalings.} 

{In principle, a comprehensive analysis of the tails could be performed. Here, we resume to simply identify the IR and UV slopes.} For both ions and electrons we identify a $5/2$ slope for the transfers towards smaller scales (UV transfers). Regarding the exchanges with the larger scales (IR transfers), we identify a $4/2$ scaling for the ions and a $3/2$ scaling for the electrons. {These scalings can be identified as the locality exponents for the energy cascade, which is contained in the full locality problem, i.e. the locality of interactions.} We try next to recover these scalings from a set of modified locality functions.

\subsection{Modified locality functions}

{We proceed now to impose the $q\ge q_{\min}$ condition for the locality functions, where we take $q_{\min} = \min\{k,p\}$. Starting from the locality functions given by the eqs.~(\ref{IRtwo}-\ref{UVtwo}) in terms of the infinitesimal triple-scale transfer $\mathcal S(k|p|q)$ and accounting for the ${q\ge\min\{k,p\}}$ condition, we define the modified locality functions as }
\begin{align}
\widetilde \Pi_{\mbox{\scriptsize ir}}\,\! (k_\kp|k_\kc)  &=  \int_{k_c}^{\infty} dk \left[\int_{0}^{k_\kp}  dp \int_{p}^{\infty}   dq\ \mathcal S(k|p|q)  \right]  \label{IRmodtwo}\ , \\
\widetilde \Pi_{\mbox{\scriptsize uv}}\,\! (k_\kp|k_\kc) &=  \int_{0}^{k_c} dp \left[\int_{k_p}^{\infty}  dk \int^{\infty}_{p}   dq\ \mathcal S(k|p|q) + \int^{\infty}_{k_\kp}  dq \int_{0}^{k_p}   dk\ \mathcal S(k|p|q)\right]  \label{UVmodtwo}\ . 
\end{align}
{In terms of our waveband decomposition, the equivalent definitions used in actual computations are given as
\begin{align}
\widetilde \Pi_{\mbox{\scriptsize ir}}\,\! (k_\kp|k_\kc) \!&=\! \! \sum_{\si=\kc+1}^{N} \! \left[\sum_{\sii=1}^{\kp} \sum_{\siii=\sii+1}^N  \right] \! \mathcal S({\si|\sii|\siii})\;, \\
\widetilde \Pi_{\mbox{\scriptsize uv}}\,\! (k_\kp|k_\kc) \!&=\! \! \sum_{\sii=1}^{\kc} \! \left[\sum_{\si=\kp+1}^{N} \sum_{\siii=\sii+1}^N \!+\! \sum_{\siii=\kp+1}^{N} \sum_{\si=1}^{\kp}  \right] \! \mathcal S({\si|\sii|\siii})\;. 
\end{align}
}

\begin{figure}[t]
\center
\includegraphics[width = 0.8\textwidth]{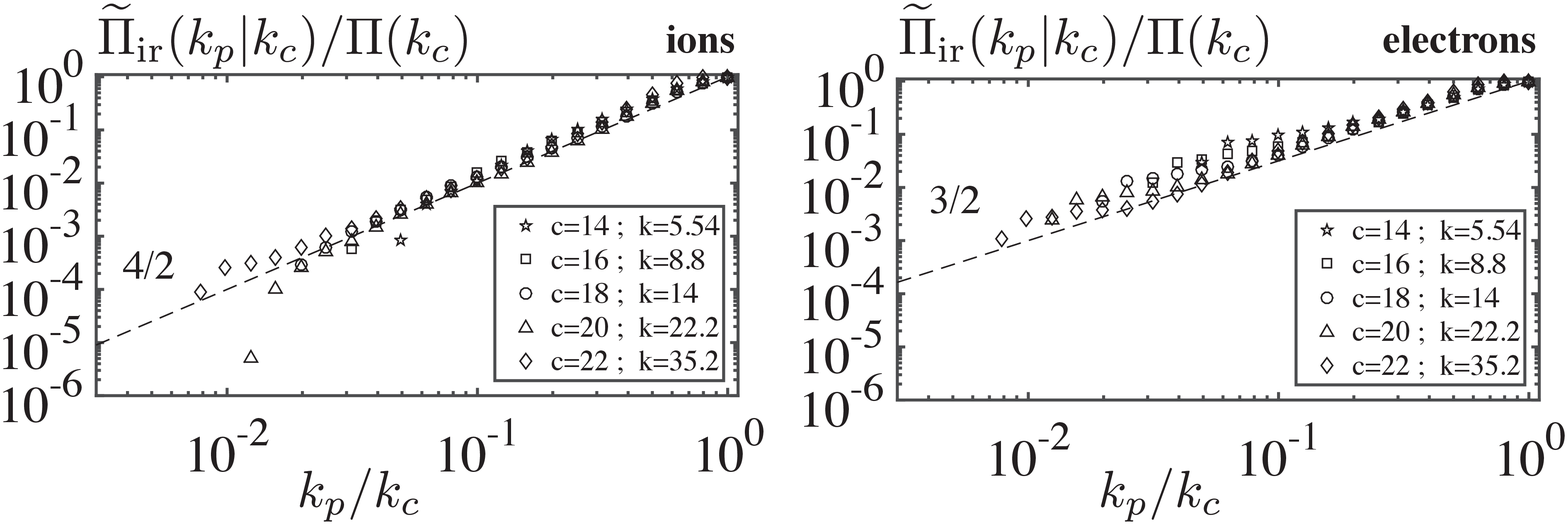}
\caption{The modified IR locality functions for the (left) ions and (right) electrons.}
\label{fig_GK-IR}
\end{figure}

\begin{figure}[t]
\center
\includegraphics[width = 0.8\textwidth]{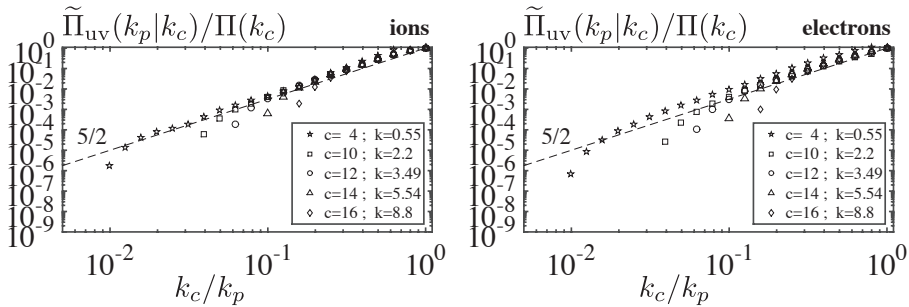}
\caption{The modified UV locality functions for the (left) ions and (right) electrons.}
\label{fig_GK-UV}
\end{figure}

We look at the IR in figure~\ref{fig_GK-IR} and UV in figure~\ref{fig_GK-UV} for the modified locality functions considered here. The locality curves show a clear power law and they collapse on each other for different cutoffs ($k_c$). This is a clear sign that we recover asymptotic locality exponents. {Furthermore, using these modified locality functions, we find a $4/2$ value for the ion's IR exponent, a $3/2$ value for the electron's IR exponent  and a $5/2$ value for both UV exponents. These are the same values as the ones found for the locality exponents of the cascade. As such, we see that we can recover these exponents directly from the scaling of the tails of the scale-to-scale transfers. However, we still need to relate these asymptotic locality exponents (for the energy cascade) with the ones in section~\ref{sec:loc} (the exponents for the locality of interaction). We do so next and make our final conclusions.}

\section{Discussion and conclusions}  \label{sec:conclusions}

\subsection{Discussing the relation between the IR locality function and its modified form}

{We first mention that in the UV limit, the locality exponents for the nonlinear interactions (obtained via the normal locality functions) and the asymptotic locality exponents found for the energy cascade (either directly from the scale-to-scale transfers or via the use of the modified locality functions) recover the same $5/2$ value for both ions and electrons.}

\begin{figure}[t]
\center
\includegraphics[width = 0.8\textwidth]{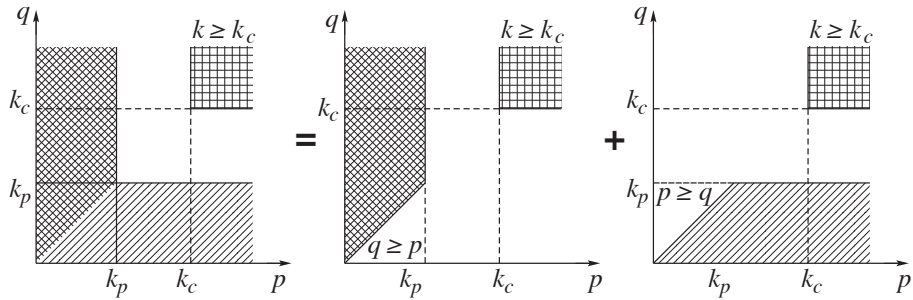}
\caption{The geometric relation between the two terms entering the IR locality function, eq.~(\ref{IRtwo}), for known $k_p$ and $k_c$ values. The first term constitutes the modified IR locality function.}
\label{fig_IRgeometry}
\end{figure}

Compared to the UV case, the IR locality exponents obtained for the full and modified definitions {of the locality functions} differ drastically. To understand better the significance behind this difference, we plot in figure \ref{fig_IRgeometry} the ($p, q$) integration domain of $\mathcal S(k|p|q)$ for eq.~(\ref{IRtwo}) and its two constituent terms. Since in the $\mathcal S(k|p|q)$ object the position of the $k,p,q$ scales matters, the two terms in eq.~(\ref{IRtwo}) have different physical interpretations.

Due to the definition of IR locality functions, the ordering $k \ge k_c \ge k_p$ is always valid. Considering now solely the $q\ge p$ (first) term, for which $k_p \ge p$, we see from the condition listed in eq.~(\ref{qmin}) {(or equivalently in eq.~\ref{qlim})} that we always have {at least a} dyadic separation between the giving and receiving scales (i.e.  $k\ge 2p$) for $k_c \ge 2k_p$. As explained in {section \ref{sec:dyad}}, such interactions are mediated only by immediate dyadic structures above and below the {energy} receiving scales. {Thus, these interactions are quite local and as a consequence scale very well with the {energy} receiving scale {$k$}, i.e. the dynamics of these interactions tend to become scale invariant and recover asymptotic locality properties.} The separation imposed by $k_p$ and $k_c$ limits now the separation between the giving and receiving scales and the locality exponents thus measured are directly related to the locality of the energy cascade. The modified IR locality functions stated in eq.~(\ref{IRmodtwo}) is simply this first term, which measures the intensity of the energy cascade as a function of the separation between the energy exchanging scales and leads to the recovery of asymptotic locality exponents. {We mentions that for $k_c=k_p$ the $q\ge p$ conditions can select intra-dyadic exchanges, however, these exchanges tend to be mediated by $q\sim k$ and do not increase the nonlocal nature of interactions. Moreover, interactions of the type $q=p=k$ cannot be seen as moving energy from one scale to another.}

In the second term, we have $p\ge q$ and $k \ge k_c \ge  k_p \ge q$. This limitation on the mediator scale $q$ selects only transfers of energy between scales contained in the same dyadic signal. While the {energy exchanges are indeed very local, the largest contributions involve couplings for which the mediator is well separated} from the energy exchanging scales $q \ll p\sim k$. This leads to a strong nonlocal character for the nonlinear interactions, {which} is captured by the second term's exponents. The fact that the mediation of {intra-dyadic} transfers $p\sim k$ is done by scales comparable to the forcing range, scales which are thus amplified directly by the linear forcing term, is listed as a warning to the impact made by a force on the development of turbulence \cite{Aluie:2010p946}.  

The full IR locality function captures both of these distinct aspects and, depending on their intensity, provides an effective locality exponent. {As we have shown, asking for $k$ and $p$ to be at least dyadic separated leads to a null contribution for the second term.} This fact emphasises the interpretation that the second term measures the locality of plane waves interacting inside the same dyadic signal and its not a measure of the classical turbulence energy cascade (overall, the energy exchange between a dyadic scale and itself is zero). Indeed, for hydrodynamics and MHD turbulence, the contribution of the second term is sufficiently reduced as to recover globally the asymptotic locality exponents. This is not the case for GK turbulence, where the exchange between neighboring waves seems to remain strong even at the smallest scales.

\subsection{Discussing the asymptotic locality exponents}

For GK turbulence, the effective IR locality exponents do show that turbulence is strongly nonlocal. However, at the same time, our detailed analysis shows that the energy exchange between well separated scale structures recovers asymptotic values for the locality exponents. In the past \cite{Teaca:2012p1415}, a value of $(k_p/k_c)^{\pm 5/6}$ was estimated for the IR and UV asymptotic locality exponents. Those values were determined using scalings computed for {statistically} homogenous two-dimensional GK turbulence \cite{Plunk:2010p1360}. Here, we do not confirm {those} predictions, as we find, respectively, $4/2$ and $3/2$ values for the ion and electron IR asymptotic locality exponents. For the UV, both ions and electrons converge on the $5/2$ value for the asymptotic locality exponent, which are recovered from the full form of the locality functions, denoting their robustness. 

The recovery of asymptotic locality exponents is indeed impressive, as it shows that in spite of all existing complications, GK turbulence possesses a strong classical characteristic for the kinetic Alfv\'en wave cascade. 
{By filtering the large scales mediators and limiting the contributions of intra-dyadic exchanges we allow the nature of the remainder interactions to surface. We see that embedded in the full GK turbulence problem, there exists an asymptotic turbulence component (just like hydrodynamical or MHD turbulence), which under ideal conditions may yet be realised. Even if only a tendency, the asymptotic locality nature of GK turbulence can still be measured and the least expensive way is through the scaling of the scale-to-scale transfers.} In the future, measuring these asymptotic locality exponents for different plasma parameters in simple magnetic configurations will tell us if this classical characteristic is also universal for GK turbulence.

{For GK turbulence to recover a general asymptotic behavior, i.e. for its nonlinear dynamics to remain scale invariant with the increase of the interval of excited scales, the contribution of intra-dyadic exchanges (mediated by large scales) should tend towards zero. In this scenario, we would recover the asymptotic locality exponent directly from the locality functions (as the second term, $p\ge q$, would become sub-dominant) and the locality of the cascade and the locality of the nonlinear interactions would become the same. If furthermore we find the asymptotic locality exponent to be unique, we could say that GK turbulence has a universal character. However, even if the intra-dyadic exchanges will always dominate the GK system, the values of the asymptotic locality exponent can still be determined from the energy cascade and their uniqueness be assessed for various plasma parameters.} 

{Last, we mention that all theoretical estimates that assume an infinite inertial range in turbulence will automatically assume that the large scales mediated intra-dyadic exchanges are zero (or infinitely small), since the largest scales that couple in a nonlocal way will always be removed by the infinite range limit (also seen as: the large shearing flows are removed and homogeneity is restored for turbulence). This last statement can be interpreted as a warning, not to over-rely on simple scaling laws for turbulence in complex systems.}

\subsection{Conclusions}  

Using a large resolution simulation of GK turbulence in slab magnetic geometry, we have analysed the energy {cascade} and the locality of interactions. {While the interactions can be deemed as being nonlocal ($1/3$ scaling for electrons and $1/12$ for the ions in the IR limit), we have shown that embedded in the full GK problem, there exists a set of couplings that tend to develop an asymptotic turbulence behavior. These couplings possess locality exponents that can be recovered directly from measurements on the energy cascade. The energy cascade was shown to be local in nature ($3/2$ scaling for electrons and $4/2$ for the ions in the IR limit) and, more importantly, it was shown that it recovers asymptotic values for the locality exponents. This may prove to be useful in the development of sub-grid scale models for GK turbulence.}

In addition, clarifying the diagnostics that can capture the asymptotic exponents for GK turbulence is important as a wide parameter space needs to be explored to evaluate the universality of turbulence at kinetic scales. Being able to extract the asymptotic locality exponents from the less expensive scale-to-scale transfers can prove to be invaluable, especially when seeking more robust time averaged results. 

From the perspective of the five-dimensional dynamics, understanding that the exchange of energy between perpendicular scales occurs in an asymptotically local way is important when seeking to understand the balance between nonlinear phase mixing (that includes the perpendicular cascade)  and linear phase mixing (Landau damping) using turbulence scaling arguments. This is left for future work.

\section*{Acknowledgements}
BT acknowledges the discussion with G. Plunk on the subject of magnetic coordinates and phase space interactions in GK turbulence. The authors would like to acknowledge the anonymous referees for their detailed comments, which led to a much better presentation of the original ideas and to an overall improvement of the manuscript. We also acknowledge the Max-Planck Princeton Center for Plasma Physics for facilitating the discussions that led to this paper. The research leading to these results has received funding from the European Research Council under the European Union's Seventh Framework Programme (FP7/2007V2013)/ERC Grant Agreement No. 277870. The Max Planck Computing and Data Facility (MPCDF) is gratefully acknowledged for providing computational resources used for this study. The gyrokinetic simulations presented in this work used additional resources of the National Energy Research Scientific Computing Center, a DOE Office of Science User Facility supported by the Office of Science of the U.S. Department of Energy under Contract No. DE-AC02-05CH11231.

\section*{References}

\providecommand{\newblock}{}

\end{document}